\documentclass[letterpaper, 10 pt, conference]{ieeeconf} 

\IEEEoverridecommandlockouts 
\usepackage{graphicx}

\usepackage[hidelinks]{hyperref}

\usepackage{amsmath}

\usepackage{float}

\usepackage{subfigure}
\usepackage{tikz}
\usepackage{tikz-3dplot}
\usetikzlibrary{patterns}

\usepackage{cuted}

\title{\LARGE \bf The Cubli: Modeling Utilizing Quaternions}

\author{Fabio Bobrow, Bruno A. Angelico and Flavius P. R. Martins
\thanks{This work was supported by the S\~ao Paulo Research Foundation (FAPESP) for the grant 2017/22130-4.}
\thanks{The authors are with the Department of Telecommunications and Control Engineer, Escola Polit\'ecnica da USP, S\~ao Paulo, SP, Brazil. The contact author is Fabio Bobrow, e-mail: fbob@usp.br}}

\begin{document}

    \maketitle
    \thispagestyle{empty}
    \pagestyle{empty}

    \begin{abstract}
        This paper performs the modeling of a Cubli, a cube with three reaction wheels mounted on orthogonal faces that becomes a reaction wheel based 3D inverted pendulum when positioned in one of its vertices. The approach novelty is that quaternions are used instead of Euler angles. One nice advantage of quaternions, besides the usual arguments to avoid singularities and trigonometric functions, is that it allows working out quite complex dynamic equations completely by hand utilizing vector notation. Modeling is performed utilizing Lagrange equations and it is validated through computer simulations and Poinsot trajectories analysis.
    \end{abstract}

    \section{Introduction}  
        
        Inverted pendulum systems have been a popular demonstration of using feedback control to stabilize open-loop unstable systems. Introduced back in 1908 by Stephenson \cite{stephenson_1908_stability}, the first solution to this problem was presented only in 1960 with Roberge \cite{roberge_1960_mechanical} and it is still widely used to test, demonstrate and benchmark new control concepts and theories \cite{guaracy_2017_robust}.
        
\begin{figure}[H]
                \centerline{\includegraphics[width=0.3\textwidth]{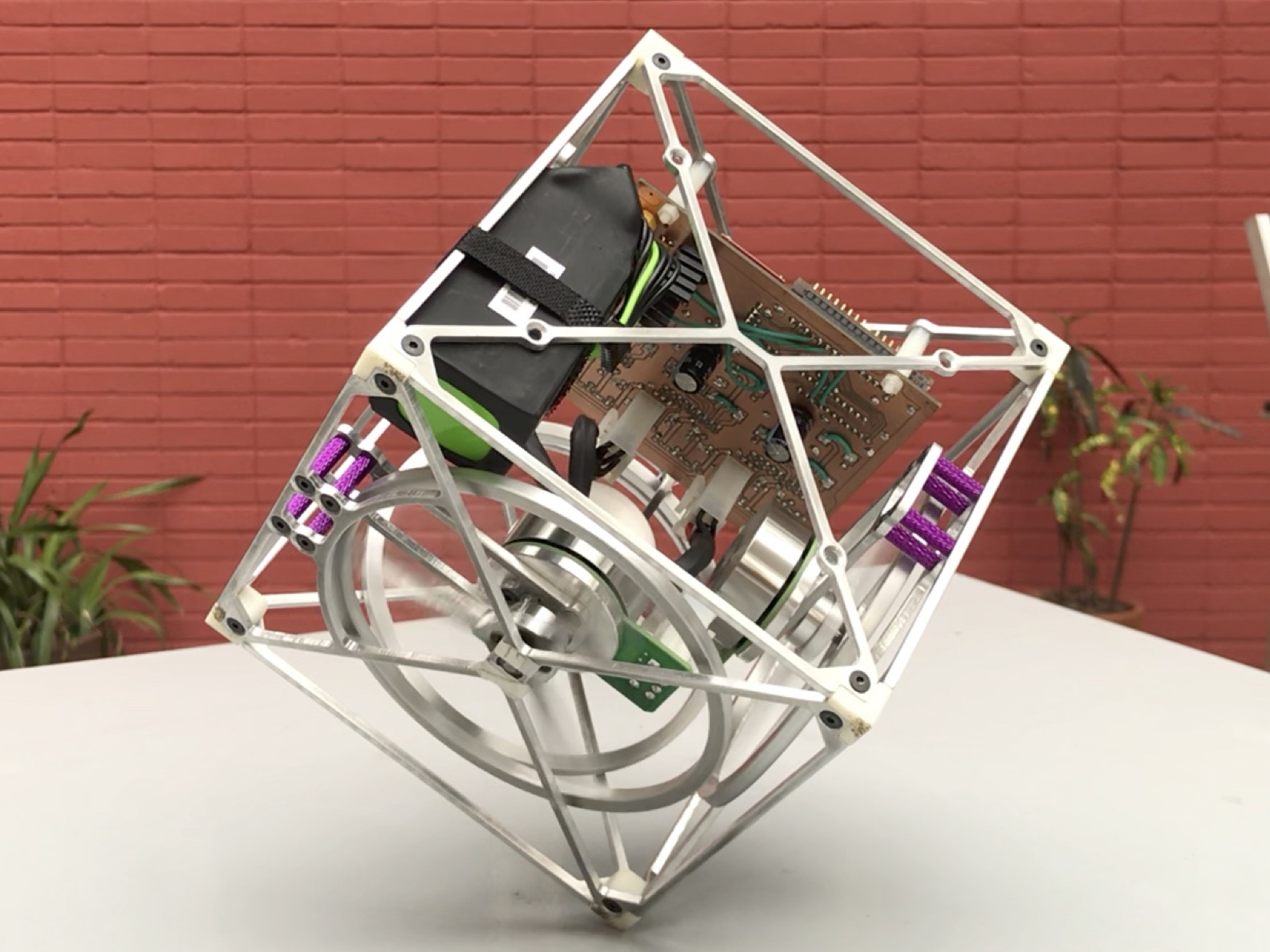}}
                \caption{Cubli} 
                \label{fig/cubli}
            \end{figure}

        Reaction wheel pendulums have a controlled rotating wheel that exchanges angular momentum with the pendulum. First introduced in 2001 by Spong \cite{spong_2001_reaction}, it was soon adapted to 3D design variants \cite{sanyal_2004_dynamics}. Perhaps, the most notable of them is Cubli (Fig. \ref{fig/cubli}), originally developed and baptized in 2012 by Gajamohan \cite{gajamohan_2012_cubli,gajamohan_2013_cubli_1} from the Institute for Dynamic Systems and Control of Zurich Federal Institute of Technology (ETH Zurich).
        
        Cubli is a device that consist of a cube with three reaction wheels mounted on orthogonal faces. By positioning the cube supported by only one of its vertices it becomes a reaction wheel based 3D inverted pendulum. This method of utilizing reaction wheels is similar to the one used for decades to stabilize satellites in space \cite{silva_2014_framework}, but due to gravity the systems dynamics are different.
        
        The purpose of this paper is to model this system. Although the ETH team has already done this and even designed and implemented a nonlinear controller \cite{gajamohan_2013_cubli_2,muehlebach_2017_cubli}, the novelty of this work will be the use of quaternions. One nice advantage of quaternions, besides the usual arguments to avoid singularities and trigonometric functions, is that it allows working out quite complex dynamic equations completely by hand utilizing vector notation \cite{graf_2008_quaternions}.
        
 \section{Modeling}
    
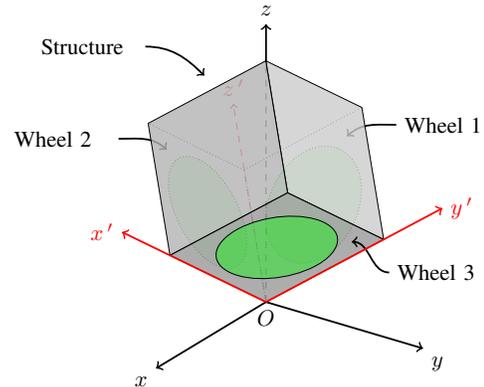
\begin{figure}[H]
                \centerline{ \scalebox{.85}{% Set angle of vision
\tdplotsetmaincoords{65}{125}

\begin{tikzpicture}[tdplot_main_coords,scale=0.6]

	% Draw inertial reference frame
	\coordinate (O) at (0,0,0) node[anchor=north]{$O$};
	\draw[->,thick] (0,0,0) -- (5,0,0) node[below left]{$x$};
	\draw[->,thick] (0,0,0) -- (0,5,0) node[below right]{$y$};
	\draw[dashed] (0,0,0) -- (0,0,6.93);
	\draw[->,thick] (0,0,6.93) -- (0,0,8) node[above]{$z$};

	% Draw body reference frame of the back
    \tdplotsetrotatedcoords{-135}{55}{135}
	\draw[tdplot_rotated_coords,->,dashed,red] (0,0,0) -- (0,0,6) node[above]{$z\,'$};
    
    % Draw reaction wheels of the back 
    \tdplotsetrotatedcoords{-15}{55}{135}
    \draw[tdplot_rotated_coords,densely dotted,fill=green,fill opacity=0.2] (2,2,0) circle(1.6);
    \draw[tdplot_rotated_coords,thick,->] (5,3,0) node[right,text width=1.5cm] 
        {Wheel 1} to [out=180,in=45] (3.4,3.4,0);
    \tdplotsetrotatedcoords{105}{55}{135}
    \draw[tdplot_rotated_coords,densely dotted,fill=green,fill opacity=0.2] (2,2,0) circle(1.6);
    \draw[tdplot_rotated_coords,thick,->] (3,5.5,0) node[left,text width=1.5cm] 
        {Wheel 2} to [out=0,in=135] (3.4,3.4,0);
    
    % Draw cube
    \tdplotsetrotatedcoords{-135}{55}{135}
    \draw[tdplot_rotated_coords,densely dotted] (0,0,0) -- (0,0,4);
    \draw[tdplot_rotated_coords,densely dotted] (0,0,4) -- (4,0,4);
    \draw[tdplot_rotated_coords,densely dotted] (0,0,4) -- (0,4,4);
    \draw[tdplot_rotated_coords,fill=gray!40,fill opacity=0.80] (0,4,0) -- (0,4,4) -- (4,4,4) -- (4,4,0) -- (0,4,0);
    \draw[tdplot_rotated_coords,fill=gray!60,fill opacity=0.80] (4,0,0) -- (4,4,0) -- (4,4,4) -- (4,0,4) -- (4,0,0);
    \draw[tdplot_rotated_coords,fill=gray!80,fill opacity=0.80] (0,0,0) -- (4,0,0) -- (4,4,0) -- (0,4,0) -- (0,0,0);
    \draw[tdplot_rotated_coords,thick,->] (5,1,5.5) node[left,text width=1.5cm] 
        {Structure} to [out=0,in=135] (4,2,4.5);
    
   	% Draw reaction wheels of the front
    \tdplotsetrotatedcoords{-135}{55}{135} 
    \draw[tdplot_rotated_coords,fill=green,fill opacity=0.4] (2,2,0) circle(1.6);
    \draw[tdplot_rotated_coords,thick,->] (0,4,-1) node[right,text width=1.5cm] 
        {Wheel 3} to [out=180,in=0] (0.6,3.4,0);

	% Draw body reference frame of the front
    \tdplotsetrotatedcoords{-135}{55}{135}
	\draw[tdplot_rotated_coords,->,thick,red] (0,0,0) -- (6,0,0) node[left]{$x\,'$};
	\draw[tdplot_rotated_coords,->,thick,red] (0,0,0) -- (0,6,0) node[right]{$y\,'$};

\end{tikzpicture}}}
                \caption{Cubli body parts} \label{fig/cubli_body_parts}
            \end{figure}

        Cubli is composed of four rigid bodies: a structure and three reaction wheels (Fig. \ref{fig/cubli_body_parts}). The structure rotates freely around the pivot point $O$ (articulation vertex), while each reaction wheel, besides rotating together with the structure, also rotates around its axial axis. 
            
        There are other bodies, such as motors, batteries, microcontrollers, etc., that can be interpreted as being part of one of them. The only exception are the motors, whereby their stators are considered part of the structure while their rotors are considered part of the reaction wheels. 
      
        \subsection{Kinematics}
        \label{sec/kinematics}

            Since the angular displacements of the reaction wheels are not important for the system dynamics, only the structure orientation will be considered in the model. Unlike Euler angles, based on the property that any orientation of a rigid body can be described with a sequence of three rotations around predefined axis, in the quaternions notation any orientation of a rigid body can be described with a single rotation around the real eigenvector of the transformation matrix between body axes and inertial axes. For this, quaternions require four parameters: three to describe the eigenaxis $\hat{e}$ coordinates plus one to describe rotation angle $\phi$. 
            
            \subsubsection{Spacial rotation}
            
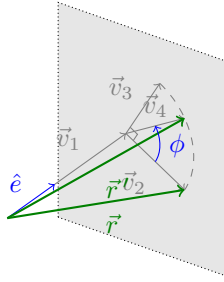
\begin{figure}[H]
                    \centerline{% Set angle of vision
\tdplotsetmaincoords{60}{35}

\begin{tikzpicture}[tdplot_main_coords,scale=0.55]

	% Draw inertial reference frame
	%\draw[thick,->] (0,0,0) -- (1,0,0);
	%\draw[thick,->] (0,0,0) -- (0,1,0);
	%\draw[thick,->] (0,0,0) -- (0,0,1);

    \draw[fill=gray!20,densely dotted] (-2,5,-3) -- (-2,5,3) -- (3,5,3) -- (3,5,-3) -- cycle;
	
    % Draw rotation axis
	\draw[->,blue] (0,0,0) -- node[above left]{$\hat{e}$} (0,2,0);
	\draw[->,gray] (0,2,0) -- node[above left]{$\vec{v}_1$} (0,5,0);
	\draw[->,gray] (0,5,0) -- node[below left]{$\vec{v}_2$} (1.732,5,-1);
	\draw[->,gray] (0,5,0) -- node[above left]{$\vec{v}_3$} (1,5,1.732);
	\draw[->,gray] (0,5,0) -- node[above]{$\vec{v}_4$} (1.732,5,1);
	\draw[->,thick,green!50!black] (0,0,0) -- node[below right]{$\vec{r}$} (1.732,5,-1);
	\draw[->,thick,green!50!black] (0,0,0) -- node[below right]{$\vec{r}\,'$} (1.732,5,1);

	\draw[gray] (0.216,5,-0.125) -- (0.341,5,0.091) -- (0.125,5,0.216);

	\tdplotsetrotatedcoords{90}{90}{0}	
	\tdplotdrawarc[<-,blue,tdplot_rotated_coords]{(0,0,5)}{1}{-120}{-60}{right}{$\phi$};
	\tdplotdrawarc[dashed,gray,tdplot_rotated_coords]{(0,0,5)}{2}{-150}{-60}{}{};
        
\end{tikzpicture}}
                	\caption{Rodrigues' rotation geometry} \label{fig/rodrigues_rotation_geometry}
                \end{figure}

                Let $\vec{r}$ be an arbitrary vector to be rotated around a unitary vector $\hat{e}$ by an angle $\phi$ generating a rotated vector $\vec{r}\,'$ (Fig. \ref{fig/rodrigues_rotation_geometry}).
                
                Projection vectors $\vec{v}_1$, $\vec{v}_2$, $\vec{v}_3$ and $\vec{v}_4$ can be written in terms of vector $\vec{r}$, unitary vector $\hat{e}$ and angle $\phi$:
                \begin{equation}
                    \left\{
                    \begin{array}{l}
                        \vec{v}_1 = (\vec{r}\cdot\hat{e})\hat{e} \\
                        \vec{v}_2 = \vec{r} - \vec{v}_1 \\
                        \vec{v}_3 = \vec{v}_2 \times \hat{e} \\
                        \vec{v}_4 = \vec{v}_2 \cos \phi + \vec{v}_3 \sin \phi
                    \end{array}
                    \right.
                    \label{eqn/projection_vectors}
                \end{equation}
                
                The rotated vector $\vec{r}\,'$ can be written in terms of projection vectors, such that $\vec{r}\,' = \vec{v}_1 + \vec{v}_4$. By using (\ref{eqn/projection_vectors}), it results in:
                \begin{equation}
                    \vec{r}\,' = (1-\cos\phi)(\vec{x}\cdot\hat{e})\hat{e} + \cos\phi\vec{r}  + \sin\phi (\vec{r} \times \hat{e}) \label{eqn/rodrigues_rotation_formula}
                \end{equation}
                
                Equation (\ref{eqn/rodrigues_rotation_formula}) is the famous Rodrigues' rotation formula that describes the rotation of a vector $\vec{r}$ by an angle $\phi$ along a unitary vector $\hat{e}$.
                
            \subsubsection{Quaternion fundamentals}
            
                Quaternion algebra can be generated from the property $i^2 = j^2 = k^2 = ijk = -1$, from where, the following multiplication rules arise: $ij = k$, $jk = i$, $ki = j$, $ji = -k$, $kj = -i$, $ik = -j$. As can be seen, the product is non-commutative.
            
            \subsubsection{Quaternion notation}
            
            	A quaternion $q$ is a set of four parameters, a real value $q_0$ and three imaginary values $q_1i$, $q_2j$ and $q_3k$:
            	\begin{equation}
            		q = q_0 + q_1i + q_2j + q_3k
            	\end{equation}
                
                A quaternion can also be represented as a four dimension column vector composed of a real value $q_0$ and a vectorial imaginary value $\vec{q} = \begin{bmatrix} q_1 & q_2 & q_3\end{bmatrix}^T$:
                \begin{equation}
                	q =
                	\begin{bmatrix}
                		q_0 & \vec{q}^T
                	\end{bmatrix}^T
                \end{equation}
                whose conjugate is represented as $
                	\bar{q} = \begin{bmatrix}
                		q_0 & -\vec{q}^T
                	\end{bmatrix}^T$, and its norm (a nonnegative real value) as:
                \begin{equation}
                	|q| = \sqrt{q_0^2+q_1^2+q_2^2+q_3^2}
                \end{equation}
            
            \subsubsection{Quaternion product}
                
                Considering the aforementioned properties, the product of two quaternions $q$ and $r$ (represented by the $\circ$ operator) can be derived:
                \begin{equation}
                    q \circ r = 
                    \begin{bmatrix}
                        q_0r_0 - \vec{q}\cdot\vec{r} \quad {\left(q_0\vec{r} + r_0\vec{q} + \vec{q}\times\vec{r}\right)}^T
                    \end{bmatrix}^T
                    \label{eqn/quaternion_product}
                \end{equation}
                
                From Eq. (\ref{eqn/quaternion_product}) it turns out that:
                \begin{equation}
                    q \circ \bar{q} = \bar{q} \circ q =
                    \begin{bmatrix}
                        |q|^2 & \vec{0}^T
                    \end{bmatrix}^T
                    \label{eqn/quaternion_property_1}
                \end{equation}
                
                \noindent
                and if the quaternion has unitary norm ($|q|=1$):
                \begin{equation}
                    q \circ \bar{q} = \bar{q} \circ q =
                    \begin{bmatrix}
                        1 & \vec{0}^T
                    \end{bmatrix}^T
                    \label{eqn/quaternion_property_2}
                \end{equation}
                
            \subsubsection{Rotation quaternion}
                
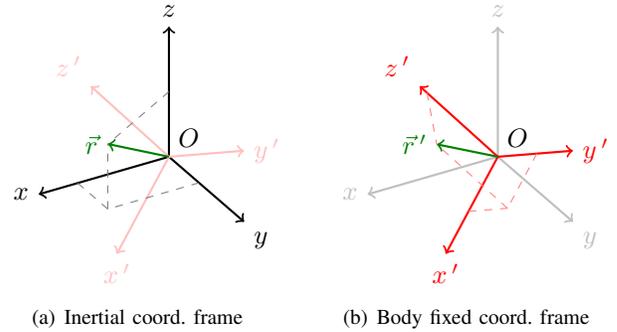
\begin{figure}[H]
                    \subfigure[][Inertial coord. frame]{% Set angle of vision
\tdplotsetmaincoords{60}{150}

\begin{tikzpicture}[tdplot_main_coords]

	% Draw inertial reference frame
	\coordinate (O) at (0,0,0) node[above right]{$O$};
	\draw[thick,->] (0,0,0) -- (2,0,0) node[left]{$x$};
	\draw[thick,->] (0,0,0) -- (0,2,0) node[below right]{$y$};
	\draw[thick,->] (0,0,0) -- (0,0,2) node[above]{$z$};
    
    % Draw rotated reference frame
	\tdplotsetrotatedcoords{-19.11}{41.41}{40.89}	
	\draw[thick,->,red!25,tdplot_rotated_coords] (0,0,0) -- (1.6,0,0) node[below]{$x\,'$};
	\draw[thick,->,red!25,tdplot_rotated_coords] (0,0,0) -- (0,1.6,0) node[right]{$y\,'$};
	\draw[thick,->,red!25,tdplot_rotated_coords] (0,0,0) -- (0,0,1.6) node[above left]{$z\,'$};
    
    % Draw r vector
	\draw[thick,->,green!50!black] (0,0,0) -- (1.4,0.8,1) node[left]{$\vec{r}$};
	\draw[dashed,gray] (1.4,0,0) -- (1.4,0.8,0);
	\draw[dashed,gray] (0,0.8,0) -- (1.4,0.8,0);
	\draw[dashed,gray] (1.4,0.8,0) -- (1.4,0.8,1);
	\draw[dashed,gray] (0,0,1) -- (1.4,0.8,1);
  
\end{tikzpicture}}
                    \subfigure[][Body fixed coord. frame]{% Set angle of vision
\tdplotsetmaincoords{60}{150}

\begin{tikzpicture}[tdplot_main_coords]

	% Draw inertial reference frame
	\coordinate (O) at (0,0,0) node[above right]{$O$};
	\draw[thick,black!25,->] (0,0,0) -- (2,0,0) node[left]{$x$};
	\draw[thick,black!25,->] (0,0,0) -- (0,2,0) node[below right]{$y$};
	\draw[thick,black!25,->] (0,0,0) -- (0,0,2) node[above]{$z$};
    
    % Draw rotated reference frame
	\tdplotsetrotatedcoords{-19.11}{41.41}{40.89}
	\draw[thick,->,red,tdplot_rotated_coords] (0,0,0) -- (1.6,0,0) node[below]{$x\,'$};
	\draw[thick,->,red,tdplot_rotated_coords] (0,0,0) -- (0,1.6,0) node[right]{$y\,'$};
	\draw[thick,->,red,tdplot_rotated_coords] (0,0,0) -- (0,0,1.6) node[above left]{$z\,'$};
    
    % Draw r vector
	\draw[thick,->,green!50!black,tdplot_rotated_coords] (0,0,0) -- (0.9,0.83,1.45) node[left]{$\vec{r}\,'$};
	\draw[dashed,red!50,tdplot_rotated_coords] (0.9,0,0) -- (0.9,0.83,0);
	\draw[dashed,red!50,tdplot_rotated_coords] (0,0.83,0) -- (0.9,0.83,0);
	\draw[dashed,red!50,tdplot_rotated_coords] (0.9,0.83,0) -- (0.9,0.83,1.45);
	\draw[dashed,red!50,tdplot_rotated_coords] (0,0,1.45) -- (0.9,0.83,1.45);
  
\end{tikzpicture}} 
                	\caption{Same vector described in different coordinate frames} \label{fig/quaternions_vector_r}
                \end{figure}

                Let $\vec{r}$ be an arbitrary fixed vector described in an inertial coordinate frame $O:\{x,y,z\}$ (Fig. \ref{fig/quaternions_vector_r}a) and $\vec{r}\,'$ be this same vector but described in a body fixed coordinate frame $O:\{x\,',y\,',z\,'\}$ (Fig. \ref{fig/quaternions_vector_r}b).
                
                To transform $\vec{r}$ into $\vec{r}\,'$, the following quaternion multiplication can be performed:
                \begin{align}
                    r\,' = \bar{q} \circ r \circ q
                    \label{eqn/quaternions_coordinate_transformation}
                \end{align}
                
                \noindent
                where $r$ and $r\,'$ are quaternions with no real part and with vectors $\vec{r}$ and $\vec{r}\,'$ in their imaginary part, i.e. $r = 
                    \begin{bmatrix}
                        0 & \vec{r}^T
                    \end{bmatrix}^T$, $r\,' = 
                    \begin{bmatrix}
                        0 & \vec{r}\,'^T
                    \end{bmatrix}^T$, and $q$ is the rotation quaternion whose components are defined in terms of the eigenaxis $\hat{e}$ and the rotation angle $\phi$, such that $
            		q =
                	\begin{bmatrix}
                		\cos\frac{\phi}{2} & \hat{e}^T\sin\frac{\phi}{2}
                	\end{bmatrix}^T
                	$. Since the eigenaxis has unitary norm ($|\hat{e}|=1$), the rotation quaternion also has unitary norm ($|q|=1$), which means that $
                    q_0^2 + q_1^2 + q_2^2 + q_3^2 = 1$.
                    
From (\ref{eqn/quaternions_coordinate_transformation}), one can see that:
                \begin{equation}
                    r\,' =
                    \begin{bmatrix}
                        0 & {\left((1-\cos\phi)(\hat{e}\cdot\vec{r})\hat{e} + \cos\phi\vec{r} + \sin\phi(\vec{r}\times\hat{e})\right)}^T
                    \end{bmatrix}^T
                    \label{eqn/quaternion_coordinate_transformation_expanded_2}
                \end{equation}
                
                \noindent
                which is identical to Rodrigues' rotation formula from (\ref{eqn/rodrigues_rotation_formula}). In order to perform the inverse transformation, one just needs to swap the rotation quaternion with its conjugate, $r = q \circ r\,' \circ \bar{q}$.
                
                Since vector $\vec{r}$ is fixed in the inertial coordinate frame, rotation quaternion $q$ can be used to represent the rotation of the body fixed coordinate frame with respect to the inertial coordinate frame.
                
            \subsubsection{Kinematic equation}
            
                Let us suppose now that the body fixed coordinate frame is in rotational motion around the origin $O$ (Fig. \ref{fig/quaternions_angular_velocity}).  Its angular velocity vector $\vec{\omega}\,'$ is given by:
        		\begin{figure}[H]
                	\centerline{% Set angle of vision
\tdplotsetmaincoords{60}{150}

\begin{tikzpicture}[tdplot_main_coords]

	% Draw inertial reference frame
	\coordinate (O) at (0,0,0) node[above right]{$O$};
	\draw[thick,->] (0,0,0) -- (2,0,0) node[left]{$x$};
	\draw[thick,->] (0,0,0) -- (0,2,0) node[below right]{$y$};
	\draw[thick,->] (0,0,0) -- (0,0,2) node[above]{$z$};
    
    % Draw rotated reference frame
	\tdplotsetrotatedcoords{-19.11}{41.41}{40.89}	
	\draw[thick,->,red,tdplot_rotated_coords] (0,0,0) -- (1.6,0,0) node[below]{$x\,'$};
	\draw[thick,->,red,tdplot_rotated_coords] (0,0,0) -- (0,1.6,0) node[right]{$y\,'$};
	\draw[thick,->,red,tdplot_rotated_coords] (0,0,0) -- (0,0,1.6) node[above left]{$z\,'$}; 
	
    % Draw angular velocities
	\tdplotsetrotatedcoords{30}{120}{90}	
	\tdplotdrawarc[->,color=blue,tdplot_rotated_coords]{(0,0,0.8)}{0.2}{-90}{210}{below right}{$\enskip\omega_x$};
	\tdplotsetrotatedcoords{-19.11}{41.41}{40.89}	
	\tdplotdrawarc[->,color=blue,tdplot_rotated_coords]{(0,0,0.8)}{0.2}{-90}{210}{below left}{$\omega_z\enskip$};
	\tdplotsetrotatedcoords{103.90}{64.34}{-146.31}	
	\tdplotdrawarc[->,color=blue,tdplot_rotated_coords]{(0,0,0.8)}{0.2}{-90}{210}{below right}{$\enskip\omega_y$};

\end{tikzpicture}}
            		\caption{Body fixed coordinate frame angular velocity} \label{fig/quaternions_angular_velocity}
            	\end{figure}
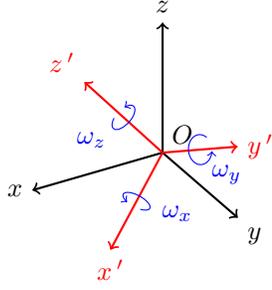

                \begin{equation}
                    \vec{\omega}\,' =
                    \begin{bmatrix}
                        \omega_x & \omega_y & \omega_z
                    \end{bmatrix}^T
                \end{equation}
                
                Note that this is the angular velocity with respect to the inertial coordinate frame but described along the body fixed coordinate frame axes.
                
                Let $\omega\,'$ be a quaternion with no real part and with the vector $\vec{\omega}\,'$ in its imaginary part, $\omega\,' =\begin{bmatrix}
                        0 & \vec{\omega}\,'^T
                    \end{bmatrix}^T$. Since vector $\vec{r}$ is fixed in the inertial coordinate frame, its time derivative as seen by the inertial coordinate frame will be zero, i.e., $
                    \dot{r} = 
                    \begin{bmatrix}
                        0 & \dot{\vec{r}}^T
                    \end{bmatrix}^T
                    =
                    \begin{bmatrix}
                        0 & \vec{0}^T
                    \end{bmatrix}^T$. On the other hand, its time derivative as seen by the body fixed coordinate frame is given by:
                \begin{equation}
                    \dot{r}\,' = 
                    \begin{bmatrix}
                        0 & \dot{\vec{r}}\,'^T
                    \end{bmatrix}^T
                    =
                    \begin{bmatrix}
                        0 & -{\left(\vec{\omega}\,' \times \vec{r}\,'\right)}^T
                    \end{bmatrix}^T
                    \label{eqn/quaternions_vector_derivative_body}
                \end{equation}
                
                The minus sign appears because if the body coordinate frame rotates in one direction, the vector will be seen by the body coordinate frame as rotating in the opposite direction. Since quaternions $r\,'$ and $\omega\,'$ have no real part, Eq. (\ref{eqn/quaternions_vector_derivative_body}) is equivalent to:
                \begin{equation}
                    \dot{r}\,' = -\omega\,' \circ r\,'
                    \label{eqn/quaternions_vector_derivative_body_2}
                \end{equation}
                
                Differentiating Eq. (\ref{eqn/quaternions_coordinate_transformation}), yields:
                \begin{equation}
                    \dot{r}\,' = - 2 \bar{q} \circ \dot{q} \circ r\,'  \label{eqn/quaternions_coordinate_transformation_derivative_2}
                \end{equation}
                
                Comparing (\ref{eqn/quaternions_coordinate_transformation_derivative_2}) with (\ref{eqn/quaternions_vector_derivative_body_2}), it is possible to obtain the angular velocity quaternion in terms of the rotation quaternion and its time derivative:
                \begin{equation}
                    \omega\,' = 2 \bar{q} \circ \dot{q} \label{eqn/quaternions_angular_velocity_from_quaternion}
                \end{equation}
                    
                Equation (\ref{eqn/quaternions_angular_velocity_from_quaternion}) can also be written in matrix notation, either with the rotation quaternion $q$ in evidence:
                \begin{equation}
                    \underbrace{
                    \begin{bmatrix}
                        \omega_x \\
                        \omega_y \\
                        \omega_z
                    \end{bmatrix}
                    }_{\vec{\omega}\,'}
                    = - 2
                    \underbrace{
                    \begin{bmatrix}
                        -\dot{q}_1 & \dot{q}_0 & \dot{q}_3 & -\dot{q}_2 \\
                        -\dot{q}_2 & -\dot{q}_3 & \dot{q}_0 & \dot{q}_1 \\
                        -\dot{q}_3 & \dot{q}_2 & -\dot{q}_1 & \dot{q}_0
                    \end{bmatrix}
                    }_{\dot{G}}
                    \underbrace{
                    \begin{bmatrix}
                        q_0 \\
                        q_1 \\
                        q_2 \\
                        q_3
                    \end{bmatrix}
                    }_{q}
                    \label{eqn/quaternions_angular_velocity_from_quaternion_matrix_1}
                \end{equation}
                 
                \noindent   
                or with its time derivative $\dot{q}$ in evidence:
                \begin{equation}
                    \underbrace{
                    \begin{bmatrix}
                        \omega_x \\
                        \omega_y \\
                        \omega_z
                    \end{bmatrix}
                    }_{\vec{\omega}\,'}
                    = 2
                    \underbrace{
                    \begin{bmatrix}
                        -q_1 & q_0 & q_3 & -q_2 \\
                        -q_2 & -q_3 & q_0 & q_1 \\
                        -q_3 & q_2 & -q_1 & q_0
                    \end{bmatrix}
                    }_{G}
                    \underbrace{
                    \begin{bmatrix}
                        \dot{q}_0 \\
                        \dot{q}_1 \\
                        \dot{q}_2 \\
                        \dot{q}_3
                    \end{bmatrix}
                    }_{\dot{q}}
                    \label{eqn/quaternions_angular_velocity_from_quaternion_matrix_2}
                \end{equation}
                
                By left-multiplying both sides of Eq. (\ref{eqn/quaternions_angular_velocity_from_quaternion}) with $q$ and using Eq. (\ref{eqn/quaternion_property_2}), it can be seen that $
                    \dot{q} = \frac{1}{2} q \circ \omega\,'$, which can also be written in matrix notation, either with the rotation quaternion $q$ in evidence:
                \begin{equation}
                    \underbrace{
                	\begin{bmatrix}
                		\dot{q}_0 \\
                        \dot{q}_1 \\
                        \dot{q}_2 \\
                        \dot{q}_3 \\
                	\end{bmatrix}
                	}_{\dot{q}}
                    =
                    \frac{1}{2}
                    \underbrace{
                    \begin{bmatrix}
                    	0 & -\omega_x & -\omega_y & -\omega_z \\
                        \omega_x & 0 & \omega_z & -\omega_y \\
                        \omega_y & -\omega_z & 0 & \omega_x \\
                        \omega_z & \omega_y & -\omega_x & 0
                    \end{bmatrix}
                    }_{\Omega}
                    \underbrace{
                	\begin{bmatrix}
                		q_0 \\
                        q_1 \\
                        q_2 \\
                        q_3 \\
                	\end{bmatrix}
                	}_{q} 
                	\label{eqn/quaternions_quaternion_from_angular_velocity_matrix_1}
                \end{equation}
                
                \noindent
                or with the angular velocity vector $\vec{\omega}\,'$ in evidence:
                \begin{equation}
                    \underbrace{
                    \begin{bmatrix}
                        \dot{q}_0 \\
                        \dot{q}_1 \\
                        \dot{q}_2 \\
                        \dot{q}_3
                    \end{bmatrix}
                	}_{\dot{q}}
                    =
                    \frac{1}{2}
                    \underbrace{
                    \begin{bmatrix}
                        -q_1 & -q_2 &  -q_3 \\
                        q_0 & -q_3 & q_2 \\
                        q_3 & q_0 & -q_1 \\
                        -q_2 & q_1 & q_0
                    \end{bmatrix}
                    }_{G^T}
                    \underbrace{
                    \begin{bmatrix}
                        \omega_x \\
                        \omega_y \\
                        \omega_z
                    \end{bmatrix}
                    }_{\vec{\omega}\,'}
                	\label{eqn/quaternions_quaternion_from_angular_velocity_matrix_2}
                \end{equation}
                    
                Equation (\ref{eqn/quaternions_quaternion_from_angular_velocity_matrix_2}) is the famous rotational kinematic equation of a rigid body utilizing quaternions.
                
            \subsubsection{Lagrange matrix properties}
                
                The matrix $G$ has been called by recent papers \cite{graf_2008_quaternions} as the Lagrange matrix and has some interesting properties. Substituting Eq. (\ref{eqn/quaternions_quaternion_from_angular_velocity_matrix_2}) into (\ref{eqn/quaternions_angular_velocity_from_quaternion_matrix_2}) allows one to verify that:
                \begin{equation}
                    GG^T = I
                    \label{eqn/quaternions_lagrange_matrix_property_1}
                \end{equation}
                
                Also from (\ref{eqn/quaternions_angular_velocity_from_quaternion_matrix_2}), the angular velocity vector cross product can be written in terms $G$ and $\dot{G}$:
                \begin{equation}
                    \resizebox{0.5\textwidth}{!}{$
                    \underbrace{
                    \begin{bmatrix}
                        0 & -\omega_z & \omega_y \\
                        \omega_z & 0 & -\omega_x \\
                        -\omega_y & \omega_x & 0
                    \end{bmatrix}
                    }_{\tilde{\omega}\,'}
                    = 2
                    \underbrace{
                    \begin{bmatrix}
                        -q_1 & q_0 & q_3 & -q_2 \\
                        -q_2 & -q_3 & q_0 & q_1 \\
                        -q_3 & q_2 & -q_1 & q_0
                    \end{bmatrix}
                    }_{G}
                    \underbrace{
                    \begin{bmatrix}
                        -\dot{q}_1 & -\dot{q}_2 &  -\dot{q}_3 \\
                        \dot{q}_0 & -\dot{q}_3 & \dot{q}_2 \\
                        \dot{q}_3 & \dot{q}_0 & -\dot{q}_1 \\
                        -\dot{q}_2 & \dot{q}_1 & \dot{q}_0
                    \end{bmatrix}
                    }_{\dot{G}^T}
                    $}
                    \label{eqn/quaternions_lagrange_matrix_property_2}
                \end{equation}
                
                \noindent
                where $\tilde{\omega}\,'$ is the angular velocity skew-symmetric matrix corresponding to its cross product:
                \begin{equation}
                    \tilde{\omega}\,' = \vec{\omega}\,' \times =
                    \begin{bmatrix}
                    	0 & -\omega_z & \omega_y \\
                        \omega_z & 0 & -\omega_x \\
                        -\omega_y & \omega_x & 0
                    \end{bmatrix}
                \end{equation}
                
                Those two properties will be very useful when dealing with kinetics.

        \subsection{Kinetics}
        \label{sec/kinetics}

            Is this section, the kinetic equations of Cubli will be derived in terms of Lagrange equations.
            
            \subsubsection{Kinetic Energy}
            	
            	Cubli total kinetic energy is the sum of the kinetic energy of each moving body:
        		\begin{equation}
                	T = T_s + \sum_{i=1}^{3} T_{wi} 
                    \label{eqn/cubli_kinetic_energy_1}
        		\end{equation}
        		
        		\noindent
        		where $T_s$ is the kinetic energy of the structure and $T_{wi}$ is the kinetic energy of the $i$-th reaction wheel.
                
                The rotational kinetic energy depends on the angular velocity and inertia tensor.  Let $\vec{\omega}_s\,'$ be the structure angular velocity vector described along the body fixed coordinate frame but with respect to the inertial coordinate frame (Fig. \ref{fig/cubli_angular_velocities}), given by:
            	\begin{equation}
                	\vec{\omega}_s\,' = 
                    \begin{bmatrix}
						\omega_x &
						\omega_y &
						\omega_z
   				    \end{bmatrix}^T
                \end{equation}
                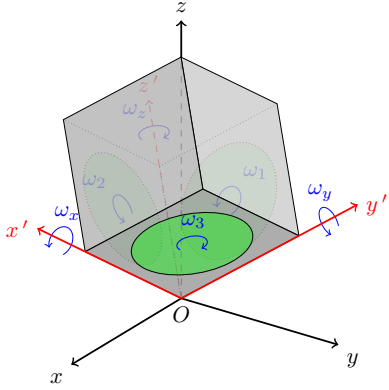
\begin{figure}[H]
                    \centerline{\scalebox{.85}{% Set angle of vision
\tdplotsetmaincoords{65}{125}

\begin{tikzpicture}[tdplot_main_coords,scale=0.6]

	% Draw inertial reference frame
	\coordinate (O) at (0,0,0) node[anchor=north]{$O$};
	\draw[->,thick] (0,0,0) -- (5,0,0) node[below left]{$x$};
	\draw[->,thick] (0,0,0) -- (0,5,0) node[below right]{$y$};
	\draw[dashed] (0,0,0) -- (0,0,6.93);
	\draw[->,thick] (0,0,6.93) -- (0,0,8) node[above]{$z$};

	% Draw body reference frame of the back
    \tdplotsetrotatedcoords{-135}{55}{135}
	\draw[tdplot_rotated_coords,->,dashed,red] (0,0,0) -- (0,0,6) node[above]{$z\,'$};
    
    % Draw reaction wheels of the back 
    \tdplotsetrotatedcoords{-15}{55}{135}
    \draw[tdplot_rotated_coords,densely dotted,fill=green,fill opacity=0.2] (2,2,0) circle(1.6);
	\tdplotdrawarc[tdplot_rotated_coords,->,blue]{(2,2,0)}{0.4}{-90}{180}{above right}{$\omega_1$}
    \tdplotsetrotatedcoords{105}{55}{135}
    \draw[tdplot_rotated_coords,densely dotted,fill=green,fill opacity=0.2] (2,2,0) circle(1.6);
	\tdplotdrawarc[tdplot_rotated_coords,->,blue]{(2,2,0)}{0.4}{-90}{180}{above left}{$\omega_2$}
    
    % Draw angular velocities of the cube
    \tdplotsetrotatedcoords{-15}{55}{135}
	\tdplotdrawarc[tdplot_rotated_coords,->,blue]{(0,0,5)}{0.4}{-90}{180}{above}{$\omega_x$}
    \tdplotsetrotatedcoords{105}{55}{135}
	\tdplotdrawarc[tdplot_rotated_coords,->,blue]{(0,0,5)}{0.4}{-90}{180}{above}{$\omega_y$}
    \tdplotsetrotatedcoords{-135}{55}{135}
	\tdplotdrawarc[tdplot_rotated_coords,->,blue]{(0,0,5)}{0.4}{-90}{180}{above left}{$\omega_z$}
    
    % Draw cube
    \tdplotsetrotatedcoords{-135}{55}{135}
    \draw[tdplot_rotated_coords,densely dotted] (0,0,0) -- (0,0,4);
    \draw[tdplot_rotated_coords,densely dotted] (0,0,4) -- (4,0,4);
    \draw[tdplot_rotated_coords,densely dotted] (0,0,4) -- (0,4,4);
    \draw[tdplot_rotated_coords,fill=gray!40,fill opacity=0.80] (0,4,0) -- (0,4,4) -- (4,4,4) -- (4,4,0) -- (0,4,0);
    \draw[tdplot_rotated_coords,fill=gray!60,fill opacity=0.80] (4,0,0) -- (4,4,0) -- (4,4,4) -- (4,0,4) -- (4,0,0);
    \draw[tdplot_rotated_coords,fill=gray!80,fill opacity=0.80] (0,0,0) -- (4,0,0) -- (4,4,0) -- (0,4,0) -- (0,0,0);
    
   	% Draw reaction wheels of the front
    \tdplotsetrotatedcoords{-135}{55}{135} 
    \draw[tdplot_rotated_coords,fill=green,fill opacity=0.4] (2,2,0) circle(1.6);
	\tdplotdrawarc[tdplot_rotated_coords,->,blue]{(2,2,0)}{0.4}{-90}{180}{above}{$\omega_3$}

	% Draw body reference frame of the front
    \tdplotsetrotatedcoords{-135}{55}{135}
	\draw[tdplot_rotated_coords,->,thick,red] (0,0,0) -- (6,0,0) node[left]{$x\,'$};
	\draw[tdplot_rotated_coords,->,thick,red] (0,0,0) -- (0,6,0) node[right]{$y\,'$};

\end{tikzpicture}}}
                        \caption{Cubli angular velocities} \label{fig/cubli_angular_velocities}
                \end{figure}
        
                Let $\vec{\omega}_{w1}\,'$, $\vec{\omega}_{w2}\,'$ and $\vec{\omega}_{w3}\,'$ be the reaction wheels relative angular velocity vectors described along and with respect to the body fixed coordinate frame (Fig. \ref{fig/cubli_angular_velocities}), given by:
                \begin{equation}
                    \vec{\omega}_{w1}\,' = \begin{bmatrix}
                        \omega_1 \\
                        0 \\
                        0
                    \end{bmatrix}, \quad
                    \vec{\omega}_{w2}\,' = \begin{bmatrix}
                        0 \\
                        \omega_2 \\
                        0
                    \end{bmatrix}, \quad
                    \vec{\omega}_{w3}\,' = \begin{bmatrix}
                        0 \\
                        0 \\
                        \omega_3
                    \end{bmatrix}
                \end{equation}
                    
                Note that this angular velocity vectors are relative; hence, to obtain the reaction wheel angular velocity vector with respect to the inertial coordinate frame, the structure angular velocity vector needs to be added (since the reaction wheels are rotating together with the structure). 
                
\begin{figure}[t!]
                    \centerline{ % Set angle of vision
\tdplotsetmaincoords{75}{115}

\begin{tikzpicture}[tdplot_main_coords,scale=0.5]

	% Draw inertial reference frame
	\coordinate[label = below:{$O$}](O) at (0,0,0);
	%\coordinate (O) at (0,0,0) node[anchor=north]{$O$};

	% Draw body reference frame of the back
    \tdplotsetrotatedcoords{0}{0}{0}
	\draw[tdplot_rotated_coords,dashed,red] (0,0,0) -- (4,0,0);
	\draw[tdplot_rotated_coords,dashed,red] (0,0,0) -- (0,4,0);
	\draw[tdplot_rotated_coords,dashed,red] (0,0,0) -- (0,0,4);
	\draw[tdplot_rotated_coords,->,thick,red] (4,0,0) -- (6,0,0) node[below left]{$x\,'$};
	\draw[tdplot_rotated_coords,->,thick,red] (0,4,0) -- (0,6,0) node[right]{$y\,'$};
	\draw[tdplot_rotated_coords,->,thick,red] (0,0,4) -- (0,0,6) node[above]{$z\,'$};
     
    % Draw lines and labels inside cube
	\coordinate[label = above right:{$G_s$}](G_c) at (2,2,2);
	\draw[tdplot_rotated_coords,->,thick,green!50!black] (0,0,0) -- (2,2,2) node[below right]{$\vec{r}_s\,'$};
    \draw[tdplot_rotated_coords,dotted,gray] (2,2,0) -- (2,2,2);
    \draw[tdplot_rotated_coords,dotted,gray] (2,2,0) -- (2,0,0);
    \draw[tdplot_rotated_coords,dotted,gray] (2,2,0) -- (0,2,0);
    \draw[tdplot_rotated_coords,dotted,gray] (0,0,2) -- (2,2,2);
       
    % Draw cube
    \tdplotsetrotatedcoords{0}{0}{0}
    \draw[tdplot_rotated_coords,fill=gray!60,fill opacity=0.40] (0,0,4) -- (4,0,4) -- (4,4,4) -- (0,4,4) -- (0,0,4);
    \draw[tdplot_rotated_coords,fill=gray!40,fill opacity=0.40] (4,0,0) -- (4,4,0) -- (4,4,4) -- (4,0,4) -- (4,0,0);
    \draw[tdplot_rotated_coords,fill=gray!80,fill opacity=0.40] (0,4,0) -- (0,4,4) -- (4,4,4) -- (4,4,0) -- (0,4,0);
    
    % Draw lines and labels outside cube
	\draw[tdplot_rotated_coords,dashed,purple!50!black] (2,2,2) -- (4,2,2);
	\draw[tdplot_rotated_coords,dashed,purple!50!black] (2,2,2) -- (2,4,2);
	\draw[tdplot_rotated_coords,dashed,purple!50!black] (2,2,2) -- (2,2,4);
    \draw[tdplot_rotated_coords,->,thick,purple!50!black] (4,2,2) -- (6,2,2)node[left]{$x''$};
    \draw[tdplot_rotated_coords,->,thick,purple!50!black] (2,4,2) -- (2,6,2) node[right]{$y''$};
    \draw[tdplot_rotated_coords,->,thick,purple!50!black] (2,2,4) -- (2,2,6) node[above]{$z''$};

\end{tikzpicture}}
                    \caption{Structure parameters}
                    \label{fig/structure_parameters}
                \end{figure}
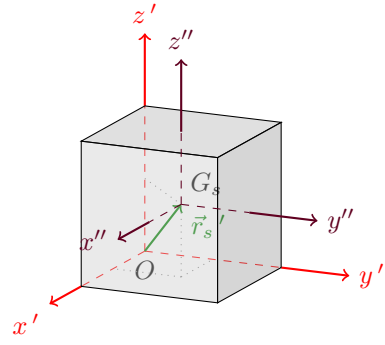

    			Let $I_{s_G}$ be the structure inertia tensor on its center of mass $G_s$ with respect to the $x''y''z''$ axes and $\vec{r}_s\,'$ be the vector from the pivot point $O$ to the structure center of mass $G_s$ (Fig. \ref{fig/structure_parameters}), given by:
                \begin{equation}
                    I_{s_G} = \text{diag}(I_{s_{xx}},I_{s_{xx}},I_{s_{xx}})
                    , \quad
                    \vec{r}_s\,' = 
                    \begin{bmatrix}
                        \frac{l}{2} & \frac{l}{2} & \frac{l}{2}
                    \end{bmatrix}^T
                \end{equation}
                
                \noindent
                where $l$ is the structure side length.
                
\begin{figure}[H]
                    \centerline{% Set angle of vision
\tdplotsetmaincoords{75}{115}

\begin{tikzpicture}[tdplot_main_coords,scale=0.5]

	% Draw inertial reference frame
	\coordinate[label = below:{$O$}](O) at (0,0,0);
	%\coordinate (O) at (0,0,0) node[anchor=north]{$O$};

	% Draw body reference frame of the back
    \tdplotsetrotatedcoords{0}{0}{0}
	\draw[tdplot_rotated_coords,->,thick,red] (0,0,0) -- (6,0,0) node[below left]{$x\,'$};
	\draw[tdplot_rotated_coords,->,thick,red] (0,0,0) -- (0,6,0) node[right]{$y\,'$};
	\draw[tdplot_rotated_coords,->,thick,red] (0,0,0) -- (0,0,6) node[above]{$z\,'$};
     
    % Draw lines and labels inside reaction wheel
	\coordinate[label = above right:{$G_{w1}$}](G_{r1}) at (0,2,2);
    \draw[tdplot_rotated_coords,thick,->,green!50!black] (0,0,0) -- (0,2,2) node[below right]{$\vec{r}_{w1}\,'$};
    \draw[tdplot_rotated_coords,dotted,gray] (0,2,0) -- (0,2,2);
    \draw[tdplot_rotated_coords,dotted,gray] (0,0,2) -- (0,2,2);
    \draw[tdplot_rotated_coords,purple!50!black,dashed] (0,2,2) -- (0,3.6,2);
    \draw[tdplot_rotated_coords,purple!50!black,dashed] (0,2,2) -- (0,2,3.6);
       
    % Draw reaction wheel
    \tdplotsetrotatedcoords{0}{-90}{0}
    \draw[tdplot_rotated_coords,fill=green,fill opacity=0.4] (2,2,0) circle(1.6);
    
    % Draw lines and labels outside reaction wheel
    \tdplotsetrotatedcoords{0}{0}{0}
    \draw[tdplot_rotated_coords,->,thick,purple!50!black] (0,2,2) -- (4,2,2) node[left]{$x_1''$};
    \draw[tdplot_rotated_coords,->,thick,purple!50!black] (0,3.6,2) -- (0,5,2) node[right]{$y_1''$};
    \draw[tdplot_rotated_coords,->,thick,purple!50!black] (0,2,3.6) -- (0,2,5) node[above]{$z_1''$};
	%\coordinate[label = left:{$J_{r,xx}$}](rx) at (4,2,2);
	%\coordinate[label = right:{$J_{r,yy}$}](ry) at (0,5,2);
	%\coordinate[label = $J_{r,yy}$](rz) at (0,2,5);

\end{tikzpicture}}
                    \caption{Reaction wheel $1$ parameters} \label{fig/reaction_wheel_parameters}
                \end{figure}
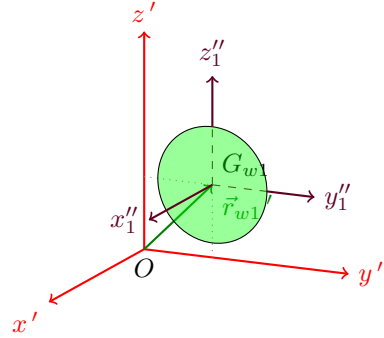

                Due to symmetry, the moment of inertia around all axes is the same. %Although some electric components may compromise this symmetry, they are being despised. 
                Let $I_{{w1}_G}$ be reaction wheel $1$ inertia tensor on its center of mass $G_{w1}$ with respect to the $x_1''y_1''z_1''$ axes and $\vec{r}_{w1}\,'$ be the vector from the pivot point $O$ to reaction wheel $1$ center of mass $G_{w1}$ (Fig. \ref{fig/reaction_wheel_parameters}),
                \begin{equation}
                    I_{{w1}_G} = \text{diag}(I_{w_{xx}},I_{w_{yy}},I_{w_{yy}})
                    , \quad
                    \vec{r}_{w1}\,' = 
                    \begin{bmatrix}
                        0 & \frac{l}{2} & \frac{l}{2}
                    \end{bmatrix}^T
                \end{equation}
                
                Due to symmetry, the moment of inertia around $y_1''$ and $z_1''$ axes are the same. Since all three reaction wheels are identical and differ only in their position, orientation and axis around which they rotate, it can be inferred that:
                \begin{align}
                    I_{{w2}_G} &= \text{diag}(I_{w_{yy}},I_{w_{xx}},I_{w_{yy}})
                    , \quad
                    \vec{r}_{w2}\,' = 
                    \begin{bmatrix}
                        \frac{l}{2} & 0 & \frac{l}{2}
                    \end{bmatrix}^T \\
                    I_{{w3}_G} &= \text{diag}(I_{w_{yy}},I_{w_{yy}},I_{w_{xx}})
                    , \quad
                    \vec{r}_{w3}\,' = 
                    \begin{bmatrix}
                        \frac{l}{2} & \frac{l}{2} & 0
                    \end{bmatrix}^T
                \end{align}
                
                With all these values, it is possible to calculate $I_{s_O}$ and $I_{{wi}_O}$, the structure and $i$-th reaction wheel inertia tensor on the pivot point $O$ with respect to the $x'y'z'$ axes, by applying the Huygens-Steiner theorem:
                \begin{align}
                    I_{s_O} &= I_{s_G} + m_s \tilde{r}_s\,' {\tilde{r}_s\,'}^T \\
                    I_{{wi}_O} &= I_{{wi}_G} + m_w \tilde{r}_{wi}\,' {\tilde{r}_{wi}\,'}^T
                \end{align}
                
                \noindent
                where $m_s$ is the mass of the structure and $m_w$ is the mass of each reaction wheel. Now it is possible to write down the total kinetic energy of Cubli from (\ref{eqn/cubli_kinetic_energy_1}):
        		\begin{align}
        			T =& \frac{1}{2} {\vec{\omega}_s\,'}^T I_{s_O} \vec{\omega}_s\,' + \sum_{i=1}^{3} \left( \frac{1}{2} {(\vec{\omega}_s\,' + \vec{\omega}_{wi}\,')}^T I_{{wi}_G} (\vec{\omega}_s\,' + \right. \nonumber \\
        			& \left.  \vec{\omega}_{wi}\,') + \frac{1}{2} {\left(\vec{\omega}_s\,' \times \vec{r}_{wi}\,'\right)}^T m_w \left(\vec{\omega}_s\,' \times \vec{r}_{wi}\,'\right) \right)
                    \label{eqn/cubli_kinetic_energy_2}
        		\end{align}
                
                Because each reaction wheel rotates around an axis orthogonal to each other, Eq. (\ref{eqn/cubli_kinetic_energy_2}) it simplified to:
            	\begin{equation}
            		T = \frac{1}{2} {\vec{\omega}_c\,'}^T \bar{I}_c \vec{\omega}_c\,' + \frac{1}{2} {(\vec{\omega}_c\,' + \vec{\omega}_w\,')}^T I_w (\vec{\omega}_c\,' + \vec{\omega}_w\,')
                    \label{eqn/cubli_kinetic_energy_3}
            	\end{equation}
                
                \noindent
                where vector $\vec{\omega}_c\,'$ is Cubli angular velocity vector, which is the same as the structure, i.e., $\vec{\omega}_c\,' = 
                    \begin{bmatrix}
                        \omega_x &
                        \omega_y &
                        \omega_z
                    \end{bmatrix}^T
                $, and vector $\vec{\omega}_w\,'$ is the composition of all three relative angular velocities vectors of the reaction wheels, $
               	 \vec{\omega}_w\,' = \begin{bmatrix}
            						\omega_1 &
                    				\omega_2 &
                    				\omega_3
            					\end{bmatrix}^T	$. Matrix $I_w$ is the net inertia tensor of the three reaction wheels around each of their individual rotational axis, $I_w = \text{diag}(I_{w_{xx}},I_{w_{xx}},I_{w_{xx}})$, 
            	and matrix $\bar{I}_c$ is Cubli total inertia tensor on pivot point $O$ only without the reaction wheels moments of inertia around each of their individual rotational axis, $
            		\bar{I}_c = \underbrace{\left(I_{s_O} + \sum_{i=1}^{3} I_{{wi}_O}\right)}_{\bar{I}_c} - I_w$.
              
		    \subsubsection{Potential Energy}
		    
                Cubli total potential energy is given by:
                \begin{equation}
                    V = V_s + \sum_{i=1}^3 V_{wi}
                    \label{eqn/cubli_potential_energy_1}
                \end{equation}
                where $V_s$ is the potential energy of the structure and $V_{wi}$ is the potential energy of the $i$-th reaction wheel. $V$ depends on the masses, center of mass vectors and acceleration of gravity vector. The first two have already been defined, remaining only the third. 
                
                Let $\vec{g}$ be the acceleration of gravity vector described in the inertial coordinate frame, $\vec{g} = \begin{bmatrix}
                        0 &
                        0 &
                        g
                    \end{bmatrix}^T$. In the body fixed coordinate frame, it is simply the rotation of the previous vector, $\vec{g}\,' = R \vec{g}$. Now it is possible to write down the total potential energy of Cubli:
                \begin{equation}
                    V = m_s {\vec{r}_s\,'}^T R \vec{g} + \sum_{i=1}^3 m_w {\vec{r}_{wi}\,'}^T R \vec{g}
                    \label{eqn/cubli_potential_energy_2}
                \end{equation}
                which can be simplified to:
                \begin{equation} 
                    V = m_c {\vec{r}_c\,'}^T R \vec{g}
                    \label{eqn/cubli_potential_energy_3}
                \end{equation}
                where $m_c$ is Cubli total mass ($m_c = m_s + 3 m_w$), and $\vec{r}_c\,'$ is the vector from the pivot point $O$ to Cubli center of mass $G_c$, i.e., $
                    \vec{r}_c\,' = \frac{m_s \vec{r}_s\,' + m_w \sum_{i=1}^3 \vec{r}_{wi}\,'}{m_s + 3 m_w}$.
              
		    \subsubsection{Lagrange equations}
		    
		        The Lagrangian, $L = T-V$, is given by:
                \begin{align}
                    L =& \frac{1}{2} {\vec{\omega}_c\,'}^T \bar{I}_c \vec{\omega}_c\,' + \frac{1}{2} {(\vec{\omega}_c\,' + \vec{\omega}_w\,')}^T I_w (\vec{\omega}_c\,' + \vec{\omega}_w\,') \nonumber \\ &- m_c {\vec{r}_c\,'}^{T} R \vec{g}
                    \label{eqn/cubli_lagrangian}
                \end{align}
                
                The kinetic equations of Cubli can then be obtained applying the Lagrange equations:
                \begin{equation}
                    \frac{d}{dt} \left( \frac{\partial L}{\partial \dot{Q}_i} \right) - \frac{\partial L}{\partial Q_i} = \sum F_{Q_i}
                 \end{equation}
                where $Q_i$ is the generalized coordinates of the system, and $F_{Q_i}$ the generalized forces in the $Q_i$ direction.
    
                There are two generalized coordinates of interest in Cubli: the structure rotation quaternion $q$ and the reaction wheels relative angular displacement vector $\vec{\theta}_w\,'$.
                
                As for the generalized forces, there is only the vector of torques $\vec{\tau}\,'$ from the motors applied on each of the three reaction wheels, given by $\vec{\tau}\,' = \begin{bmatrix} \tau_1 & \tau_2 & \tau_3  \end{bmatrix}^T$. These torques occur in the same direction as the reaction wheels relative angular displacement, which means that: $
            		\sum F_q = \vec{0}$ and $\sum F_{\vec{\theta}_w\,'} = \vec{\tau}\,'$. For each one of these generalized coordinates, there will be one kinetic equation to be calculated separately.
                	
                Applying the Lagrange equations for $Q_i = q$, the Lagrangian can be written substituting the angular velocity vector $\vec{\omega}_c\,'$ with Eq. (\ref{eqn/quaternions_angular_velocity_from_quaternion_matrix_1}) when differentiating with respect to $q$, and with Eq. (\ref{eqn/quaternions_angular_velocity_from_quaternion_matrix_2}) when differentiating with respect to $\dot{q}$. This yields the following kinetic equation:
             	\begin{align}
             		2 G^T \bar{I}_c \dot{\vec{\omega}}_c + 4 \dot{G}^T \bar{I}_c \vec{\omega}_c\,' + 2 G^T I_w \left( \dot{\vec{\omega}}_c\,' + \dot{\vec{\omega}}_w\,' \right) & \nonumber \\ + 4 \dot{G}^T I_w \left( \vec{\omega}_c\,' + \vec{\omega}_w\,' \right) + 2 m_c \Lambda  q &= \vec{0},
                    \label{eqn/cubli_kinematic_equation_structure_1}
             	\end{align}
             	where $
                	\Lambda  = \begin{bmatrix}
                		\vec{g}\,^T \vec{r}_c\,' & - (\vec{g} \times \vec{r}_c\,')^T \\
                      	- \vec{g} \times \vec{r}_c\,'  &  \vec{g}\, {\vec{r}_c\,'}^T + \vec{r}_c\,' \vec{g}\,^T - I ( \vec{g}\,^T \vec{r}_c\,' )
                	\end{bmatrix}$. Eq. (\ref{eqn/cubli_kinematic_equation_structure_1}) can be further simplified by making use of Lagrange matrix properties from (\ref{eqn/quaternions_lagrange_matrix_property_1}) and (\ref{eqn/quaternions_lagrange_matrix_property_2}):
             	\begin{align}
             		\bar{I}_c \dot{\vec{\omega}}_c\,' + \tilde{\omega}_c' \left[ \bar{I}_c \vec{\omega}_c\,' \right] + I_w \left( \dot{\vec{\omega}}_c\,' + \dot{\vec{\omega}}_w\,' \right) & \nonumber \\ + \tilde{\omega}_c' \left[ I_w \left( \vec{\omega}_c\,' + \vec{\omega}_w\,' \right) \right] + \bar{m}_c g l G \Gamma  q &= \vec{0} 
                    \label{eqn/cubli_kinematic_equation_structure_2}
             	\end{align}
             	where, $\bar{m}_c = m_s + 2m_w$ and $\Gamma  = \begin{bmatrix}
                	    1 & 1 & -1 & 0 \\
                	    1 & -1 & 0 & 1 \\
                	    -1 & 0 & -1 & 1 \\
                	    0 & 1 & 1 & 1
                	\end{bmatrix}$.
                 
                When applying the Lagrange equations for $Q_i = \vec{\theta}_w\,'$, it is also possible to rewrite the Lagrangian substituting the angular velocity vector $\vec{\omega}_w\,'$ with $\dot{\vec{\theta}}_w\,'$, which yields the following kinetic equation:
             	\begin{equation}
             		I_w \left( \dot{\vec{\omega}}_c\,' + \dot{\vec{\omega}}_w\,' \right) + \tilde{\omega}_c' \left[ I_w \left( \vec{\omega}_c\,' + \vec{\omega}_w\,' \right) \right] = \vec{\tau}\,' 
                     \label{eqn/cubli_kinematic_equation_reaction_wheels}
             	\end{equation}
                
        \subsection{Dynamics}
        \label{sec/dynamics}	
        
            Cubli kinetic equations, (\ref{eqn/cubli_kinematic_equation_structure_2}) and (\ref{eqn/cubli_kinematic_equation_reaction_wheels}), can be rewritten together in matrix notation, such as:
            \begin{equation}
                \resizebox{0.475\textwidth}{!}{$
                \begin{bmatrix}
            		\dot{\vec{\omega}}_c\,' \\
                    \dot{\vec{\omega}}_c\,' +\dot{\vec{\omega}}_w\,' \\
            	\end{bmatrix} =
                \begin{bmatrix}
                	- \bar{I}_c^{-1} \tilde{\omega}_c' \left[ \bar{I}_c \vec{\omega}_c\,' \right] - \bar{I}_c^{-1} \bar{m}_c g l G \Gamma q  \\
                	- I_w^{-1} \tilde{\omega}_c' \left[ I_w \left( \vec{\omega}_c\,' + \vec{\omega}_w\,' \right) \right]    
                \end{bmatrix} +
                \begin{bmatrix}
                	- \bar{I}_c^{-1} \\
                    I_w^{-1} 
                \end{bmatrix}
                \vec{\tau}\,'.
                $}
                \label{eqn/cubli_kinematic_equations_1}
            \end{equation}
            
            Because $\vec{\omega}_w\,' \gg \vec{\omega}_c\,'$, the kinetic equations can be simplified. Moreover, since the reaction wheels' angular velocity are relative (measured relative to the structure), their gyroscopic torques have no influence on them, only on Cubli. Hence, the reaction wheels' gyroscopic torques can be written in the first equation, rather than in the second, i.e.,
            \begin{equation}
                \resizebox{0.475\textwidth}{!}{$
                \begin{bmatrix}
            		\dot{\vec{\omega}}_c\,' \\
                    \dot{\vec{\omega}}_w\,' \\
            	\end{bmatrix} =
                \begin{bmatrix}
                	- \bar{I}_c^{-1} \tilde{\omega}_c' \left[ \bar{I}_c \vec{\omega}_c\,' + I_w \vec{\omega}_w\,' \right] - \bar{I}_c^{-1} \bar{m}_c g l G \Gamma q  \\
                	0_{3\times1}  
                \end{bmatrix} +
                \begin{bmatrix}
                	- \bar{I}_c^{-1} \\
                    I_w^{-1} 
                \end{bmatrix}
                \vec{\tau}\,'. 
                $}
                \label{eqn/cubli_kinematic_equations_2}
            \end{equation}
            
            Now, the full dynamic equations of Cubli are obtained:
            \begin{equation}
                \resizebox{0.475\textwidth}{!}{$
                \begin{bmatrix}
                	\dot{q} \\
            		\dot{\vec{\omega}}_c\,' \\
                    \dot{\vec{\theta}}_w\,' \\
                    \dot{\vec{\omega}}_w\,' \\
            	\end{bmatrix} =
                \begin{bmatrix}
                 	\frac{1}{2} G^T \vec{\omega}_c\,' \\
                	- \bar{I}_c^{-1} \left[ \tilde{\omega}_c' \left( \bar{I}_c \vec{\omega}_c\,' + I_w \vec{\omega}_w\,' \right) + \bar{m}_c g l G \Gamma q \right] \\
                	\vec{\omega}_w\,' \\
                	0_{3\times1}
                \end{bmatrix} +
                \begin{bmatrix}
                	0_{3\times3} \\
                	- \bar{I}_c^{-1} \\ 
                	0_{3\times3} \\
                    I_w^{-1} 
                \end{bmatrix}
                \vec{\tau}\,'. 
                $}
                \label{eqn/cubli_dynamic_equations}
            \end{equation}
            
            Note how compact this representation is even with the system being quite complex. It is worth mentioning that they were completely obtained by hand, without the need of any mathematical symbolic software. This was only possible because quaternions allow one to write down everything utilizing vector notation.

    \section{Validation}
    
        The model validation will be done by means of computer simulations, and they will be divided into three types: invariant analysis will be performed first, followed by singular motions and finally Poinsot trajectories. Before diving into simulations, values for all parameters of the system have to be adopted. This was done based on the CAD model of Cubli (Fig. \ref{fig/cubli}) and can be seen in Tab. \ref{tab/cubli_parameters}.
        \begin{table}[H]
            \caption{Cubli parameters}
            \label{tab/cubli_parameters}
            \centering
            \begin{tabular}{c c}
                \hline
                Parameter & Value \\
                \hline
                $l$ & $ 0.15$m \\
                $m_s$ & $0.40$ kg \\
                $m_w$ & $0.15$ kg \\
                \hline
            \end{tabular}
            \hspace{0.2cm}
            \begin{tabular}{c c}
                \hline
                Parameter & Value \\
                \hline
                $I_{s_{xx}}$ & $2\times10^{-3}$ kg.m$^2$ \\
                 $I_{w_{xx}}$ & $1\times10^{-4}$ kg.m$^2$ \\
                $I_{w_{yy}}$ & $4\times10^{-5}$ kg.m$^2$ \\
                \hline
            \end{tabular}
        \end{table}

        \subsection{Invariant analysis}
        \label{sec/invariant_analysis}
        
            It considers parameters that must remain unchanged with time when no forces are applied. Cubli has three invariants:
            \begin{itemize}
                \item Total mechanical energy $E$ 
                \item Angular momentum projection in the gravitational field direction $H_{z}$ 
                \item Angular momentum projection in the gyroscopic axis direction (diagonal axis of Cubli) $H_{z\,'}$
            \end{itemize}
            
            Its total mechanical energy is given by:
            \begin{align}
                E =& \frac{1}{2} {\vec{\omega}_c\,'}^T \bar{I}_c \vec{\omega}_c\,' + \frac{1}{2} {(\vec{\omega}_c\,' + \vec{\omega}_w\,')}^T I_w (\vec{\omega}_c\,' + \vec{\omega}_w\,') \nonumber \\ &+ m_c \vec{r}_c^{T} R \vec{g}.
                \label{eqn/cubli_mechanical_energy}
            \end{align}
            
\begin{figure}[H]
                \centerline{\includegraphics[width=0.5\textwidth]{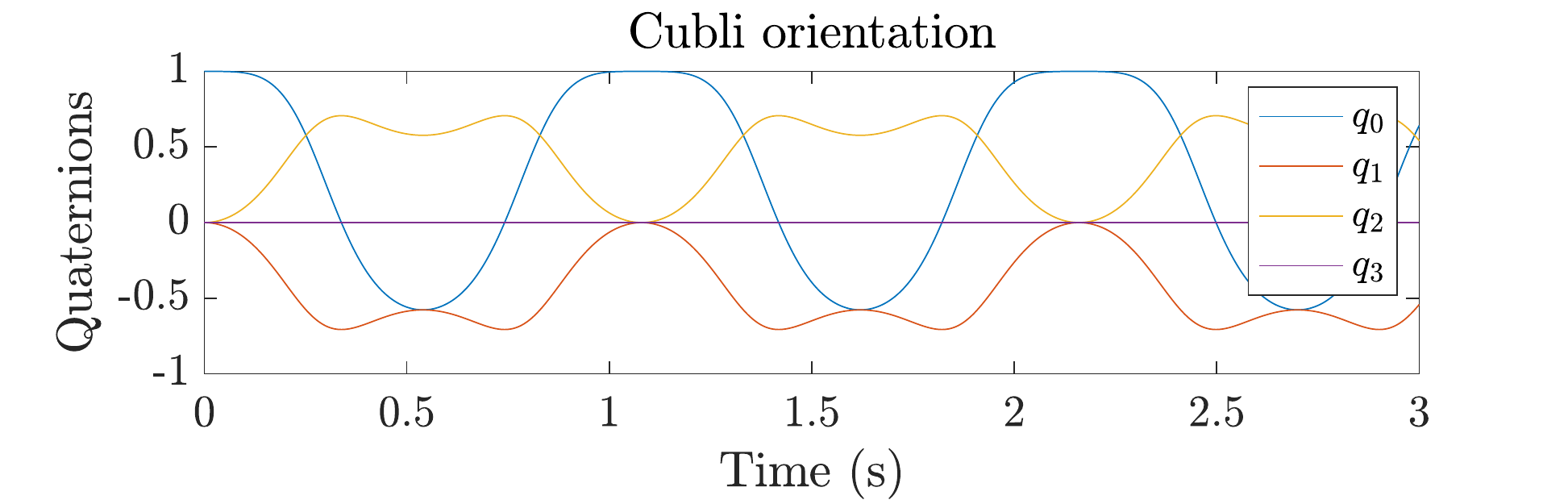}}
                \caption{Simulation 1 - Invariant analysis (quaternion)}
                \label{fig/sim1_invariant_analysis_quaternion}
            \end{figure}

            Cubli is initially aligned with the inertial reference frame, $q(0) = {\begin{bmatrix} 1 & \vec{0} \end{bmatrix}}^T$, and with $\vec{\omega}_c\,'(0) = \vec{0}$, $\vec{\omega}_w\,'(0) = \vec{0}$. Results are presented in Fig. \ref{fig/sim1_invariant_analysis_quaternion}.
            
\begin{figure}[H]
            	\centerline{\includegraphics[width=0.5\textwidth]{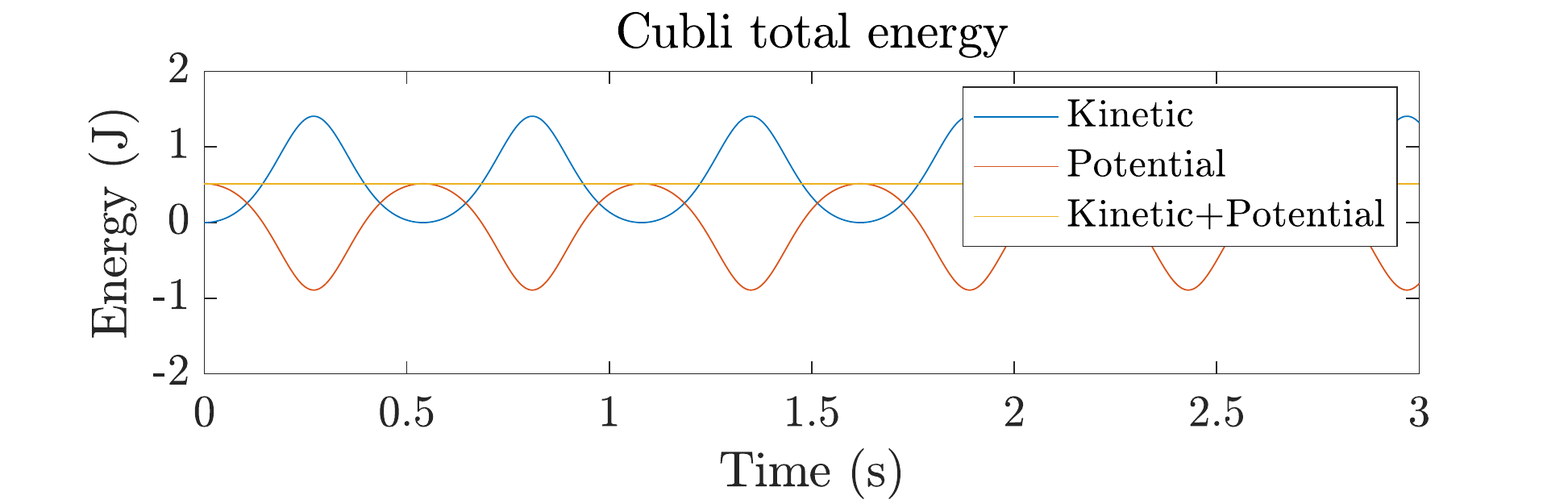}}
            	\caption{Simulation 1 - Invariant analysis (energy)}
            	\label{fig/sim1_invariant_analysis_energy}
            \end{figure}

            Since quaternions do not have an intuitive physical meaning, it is just possible to infer is that Cubli presented a kind of periodic motion. However, since the objective is to analyze the energy, this is not a problem. The mechanical energy,  presented in Fig. \ref{fig/sim1_invariant_analysis_energy}, remained unchanged. As Cubli lost potential energy, it acquired the same amount of kinetic energy, and vice-versa. This not only confirms the hypothesis of periodic motion,  but also ensures that the model is consistent.
             
       		For the angular momentum projection invariants, the reaction wheels were assumed to be fixed. The angular momentum vector is:
            \begin{equation}
                \vec{H} = \bar{I}_c \vec{\omega}_c\,',  
            \end{equation}
            so that its projections are simply given by:
            \begin{equation}
                H_z = \bar{I}_c \vec{\omega}_c\,' \cdot \vec{g}'
                , \qquad
                H_{z\,'} = \text{sum(}\bar{I}_c \vec{\omega}_c\,' \text{)}
            \end{equation}
            
\begin{figure}[H]
                \centerline{\includegraphics[width=0.5\textwidth]{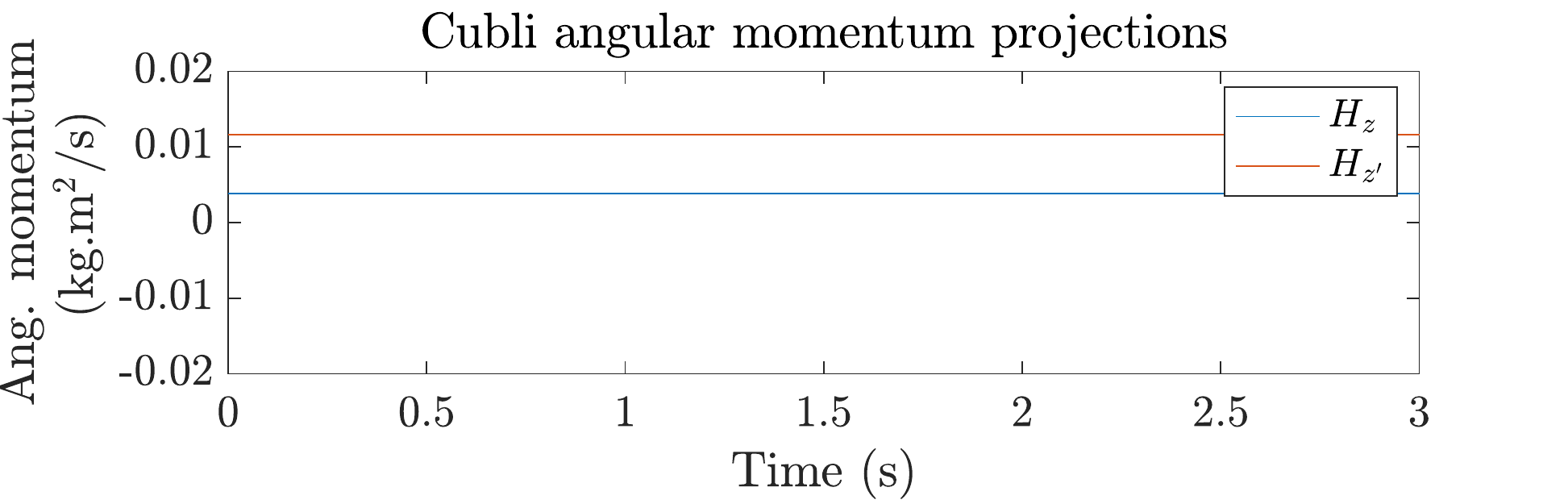}}
                \caption{Simulation 2 - Invariant analysis (angular momentum)}
                \label{fig/sim2_invariant_analysis_angular_momentum}
            \end{figure}

            Let us assume the same initial conditions as before, but now with$\vec{\omega}_c\,'(0) = \vec{1}$. As can be seen in Fig. \ref{fig/sim2_invariant_analysis_angular_momentum} the angular momentum projection on the gravitational field and on gyroscopic axis directions remained unchanged. 

        \subsection{Singular motions}
        \label{sec/singular_motions}
        
            Pre-defined initial conditions in which the behavior of the system can be predicted are considered. They will be divided into static equilibrium, whereby the system states must remain unchanged, and dynamic equilibrium, whereby the system states change as expected.
	        
            \subsubsection{Static equilibrium}
	        
\begin{figure}[H]
                    \subfigure[][Stable]{% Set angle of vision
\tdplotsetmaincoords{65}{125}

\begin{tikzpicture}[tdplot_main_coords,scale=0.425, every node/.style={scale=0.75}]

	% Draw inertial reference frame
	\coordinate (O) at (0,0,0) node[above left]{$O$};
	\draw[->,thick] (0,0,0) -- (0,5,0) node[below right]{$y$};
	\draw[->,thick] (0,0,0) -- (0,0,4) node[above]{$z$};
    
    % Draw reaction wheels of the back 
    \tdplotsetrotatedcoords{45}{125}{-45}
    \draw[tdplot_rotated_coords,densely dotted,fill=green,fill opacity=0.2] (2,2,0) circle(1.6);
    
    % Draw cube
    \tdplotsetrotatedcoords{45}{125}{-45}
    \draw[tdplot_rotated_coords,densely dotted] (4,4,0) -- (4,4,4);
    \draw[tdplot_rotated_coords,densely dotted] (0,4,0) -- (4,4,0);
    \draw[tdplot_rotated_coords,densely dotted] (4,0,0) -- (4,4,0);
    \draw[tdplot_rotated_coords,fill opacity=0.80,fill=gray!40] (0,0,0) -- (0,4,0) -- (0,4,4) -- (0,0,4) -- (0,0,0);
    \draw[tdplot_rotated_coords,fill opacity=0.80,fill=gray!60] (0,0,0) -- (0,0,4) -- (4,0,4) -- (4,0,0) -- (0,0,0);
    \draw[tdplot_rotated_coords,fill opacity=0.80,fill=gray!80] (0,0,4) -- (4,0,4) -- (4,4,4) -- (0,4,4) -- (0,0,4);
    
   	% Draw reaction wheels of the front
    \tdplotsetrotatedcoords{165}{125}{-45}
    \draw[tdplot_rotated_coords,fill=green,fill opacity=0.4] (2,2,0) circle(1.6);
    \tdplotsetrotatedcoords{-75}{125}{-45}
    \draw[tdplot_rotated_coords,fill=green,fill opacity=0.4] (2,2,0) circle(1.6);

	\draw[->,thick] (0,0,0) -- (5,0,0) node[below left]{$x$};

\end{tikzpicture}}
                    \subfigure[][Unstable]{% Set angle of vision
\tdplotsetmaincoords{65}{125}

\begin{tikzpicture}[tdplot_main_coords,scale=0.425, every node/.style={scale=0.75}]

	% Draw inertial reference frame
	\coordinate (O) at (0,0,0) node[below]{$O$};
	\draw[->,thick] (0,0,0) -- (5,0,0) node[below left]{$x$};
	\draw[->,thick] (0,0,0) -- (0,5,0) node[below right]{$y$};
	\draw[dashed] (0,0,0) -- (0,0,6.93);
	\draw[->,thick] (0,0,6.93) -- (0,0,8) node[above]{$z$};
    
    % Draw reaction wheels of the back 
    \tdplotsetrotatedcoords{-15}{55}{135}
    \draw[tdplot_rotated_coords,densely dotted,fill opacity=0.2,fill=green] (2,2,0) circle(1.6);
    \tdplotsetrotatedcoords{105}{55}{135}
    \draw[tdplot_rotated_coords,densely dotted,fill opacity=0.2,fill=green] (2,2,0) circle(1.6);
    
    % Draw cube
    \tdplotsetrotatedcoords{-135}{55}{135}
    \draw[tdplot_rotated_coords,densely dotted] (0,0,0) -- (0,0,4);
    \draw[tdplot_rotated_coords,densely dotted] (0,0,4) -- (4,0,4);
    \draw[tdplot_rotated_coords,densely dotted] (0,0,4) -- (0,4,4);
    \draw[tdplot_rotated_coords,fill opacity=0.80,fill=gray!40] (0,4,0) -- (0,4,4) -- (4,4,4) -- (4,4,0) -- (0,4,0);
    \draw[tdplot_rotated_coords,fill opacity=0.80,fill=gray!60] (4,0,0) -- (4,4,0) -- (4,4,4) -- (4,0,4) -- (4,0,0);
    \draw[tdplot_rotated_coords,fill opacity=0.80,fill=gray!80] (0,0,0) -- (4,0,0) -- (4,4,0) -- (0,4,0) -- (0,0,0);
    
   	% Draw reaction wheels of the front
    \tdplotsetrotatedcoords{-135}{55}{135} 
    \draw[tdplot_rotated_coords,fill=green,fill opacity=0.4] (2,2,0) circle(1.6);

\end{tikzpicture}} 
                    
                    \caption{Cubli static equilibrium positions} \label{fig/cubli_static_equilibrium_positions}
                \end{figure}
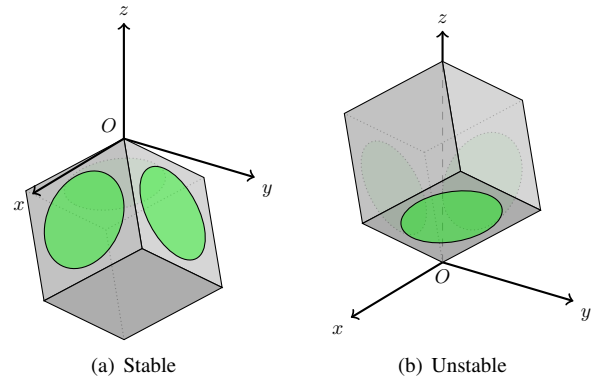

	            Cubli has two static equilibrium positions: stable and unstable, as can be seen in Fig. \ref{fig/cubli_static_equilibrium_positions}. Note that the stable one is only being considered for simulation purposes, since in reality, Cubli would never be under the floor. 

                The rotation quaternions $q_s={\begin{bmatrix} 0.46 & -0.63 & 0.63 & 0 \end{bmatrix}}^T$ and $q_u={\begin{bmatrix} 0.89 & 0.33 & -0.33 & 0 \end{bmatrix}}^T$ corresponding to the stable and unstable equilibrium positions were calculated. Since Cubli can rotate around its diagonal axis and still be in an equilibrium position, there are infinitely other equivalent rotation quaternions. In one simulation, Cubli was considered initially on its stable equilibrium position, $q(0) = q_s$, while in the other one it was considered in its unstable equilibrium position, $q(0) = q_u$. In both cases, rotation quaternions remained unchanged, confirming that these are in fact static stable positions.
                
            \subsubsection{Dynamic equilibrium}
            
\begin{figure}[H]
                    \subfigure[][Spin]{% Set angle of vision
\tdplotsetmaincoords{65}{125}

\begin{tikzpicture}[tdplot_main_coords,scale=0.425, every node/.style={scale=0.75}]

	% Draw inertial reference frame
	\coordinate (O) at (0,0,0) node[below]{$O$};
	\draw[->,thick] (0,0,0) -- (5,0,0) node[below left]{$x$};
	\draw[->,thick] (0,0,0) -- (0,5,0) node[below right]{$y$};
	\draw[dashed] (0,0,0) -- (0,0,6.93);
	\draw[->,thick] (0,0,6.93) -- (0,0,8) node[above]{$z$};
    
    % Draw reaction wheels of the back 
    \tdplotsetrotatedcoords{-15}{55}{135}
    \draw[tdplot_rotated_coords,densely dotted,fill opacity=0.2,fill=green] (2,2,0) circle(1.6);
    \tdplotsetrotatedcoords{105}{55}{135}
    \draw[tdplot_rotated_coords,densely dotted,fill opacity=0.2,fill=green] (2,2,0) circle(1.6);
    
    % Draw cube
    \tdplotsetrotatedcoords{-135}{55}{135}
    \draw[tdplot_rotated_coords,densely dotted] (0,0,0) -- (0,0,4);
    \draw[tdplot_rotated_coords,densely dotted] (0,0,4) -- (4,0,4);
    \draw[tdplot_rotated_coords,densely dotted] (0,0,4) -- (0,4,4);
    \draw[tdplot_rotated_coords,fill opacity=0.80,fill=gray!40] (0,4,0) -- (0,4,4) -- (4,4,4) -- (4,4,0) -- (0,4,0);
    \draw[tdplot_rotated_coords,fill opacity=0.80,fill=gray!60] (4,0,0) -- (4,4,0) -- (4,4,4) -- (4,0,4) -- (4,0,0);
    \draw[tdplot_rotated_coords,fill opacity=0.80,fill=gray!80] (0,0,0) -- (4,0,0) -- (4,4,0) -- (0,4,0) -- (0,0,0);
    
   	% Draw reaction wheels of the front
    \tdplotsetrotatedcoords{-135}{55}{135} 
    \draw[tdplot_rotated_coords,fill=green,fill opacity=0.4] (2,2,0) circle(1.6);
	
	% Draw angular velocities
	\tdplotdrawarc[->,blue]{(0,0,7.25)}{0.4}{-90}{180}{right}{$\quad \dot{\phi}$}

\end{tikzpicture}}
                    \subfigure[][Precession, nutation and spin]{% Set angle of vision
\tdplotsetmaincoords{65}{125}

\begin{tikzpicture}[tdplot_main_coords,scale=0.425, every node/.style={scale=0.75}]

	% Draw inertial reference frame
	\coordinate (O) at (0,0,0) node[below]{$O$};
	\draw[->,thick] (0,0,0) -- (5,0,0) node[below left]{$x$};
	\draw[->,thick] (0,0,0) -- (0,5,0) node[below right]{$y$};
	\draw[dashed] (0,0,0) -- (0,0,5);
	\draw[->,thick] (0,0,5) -- (0,0,8) node[above]{$z$};

	% Draw axis of the back
    \tdplotsetrotatedcoords{-128}{40}{116}
	\draw[tdplot_rotated_coords,dashed,gray] (0,0,0) -- (5,5,5);
    
    % Draw reaction wheels of the back 
    \tdplotsetrotatedcoords{85}{55}{160}
    \draw[tdplot_rotated_coords,densely dotted,fill opacity=0.2,fill=green] (2,2,0) circle(1.6);
    \tdplotsetrotatedcoords{162}{106}{143}
    \draw[tdplot_rotated_coords,densely dotted,fill opacity=0.2,fill=green] (2,2,0) circle(1.6);
    
    % Draw cube
    \tdplotsetrotatedcoords{-128}{40}{116}
    \draw[tdplot_rotated_coords,densely dotted] (0,0,0) -- (0,0,4);
    \draw[tdplot_rotated_coords,densely dotted] (0,0,4) -- (4,0,4);
    \draw[tdplot_rotated_coords,densely dotted] (0,0,4) -- (0,4,4);
    \draw[tdplot_rotated_coords,fill opacity=0.80,fill=gray!40] (0,4,0) -- (0,4,4) -- (4,4,4) -- (4,4,0) -- (0,4,0);
    \draw[tdplot_rotated_coords,fill opacity=0.80,fill=gray!60] (4,0,0) -- (4,4,0) -- (4,4,4) -- (4,0,4) -- (4,0,0);
    \draw[tdplot_rotated_coords,fill opacity=0.80,fill=gray!80] (0,0,0) -- (4,0,0) -- (4,4,0) -- (0,4,0) -- (0,0,0);
    
   	% Draw reaction wheels of the front
    \tdplotsetrotatedcoords{-128}{40}{116}
    \draw[tdplot_rotated_coords,fill=green,fill opacity=0.4] (2,2,0) circle(1.6);
	
	% Draw angular velocities
	\tdplotdrawarc[dashed,gray]{(0,0,6.5)}{2.4}{0}{360}{}{};
	\tdplotdrawarc[->,blue]{(0,0,6.5)}{3}{-90}{-60}{above left}{$\dot{\psi}$};
    \tdplotsetrotatedcoords{0}{20}{0};
	\tdplotdrawarc[tdplot_rotated_coords,->,blue]{(0,0,7.25)}{0.4}{-90}{180}{above right}{$\quad \dot{\phi} $};
    \tdplotsetrotatedcoords{90}{90}{0};
	\tdplotdrawarc[tdplot_rotated_coords,->,blue]{(0,0,0)}{7.5}{180}{200}{above}{$\theta$};

\end{tikzpicture}} 
                    \caption{Cubli dynamic equilibrium motions} \label{fig/cubli_dynamic_equilibrium_motions}
                \end{figure}
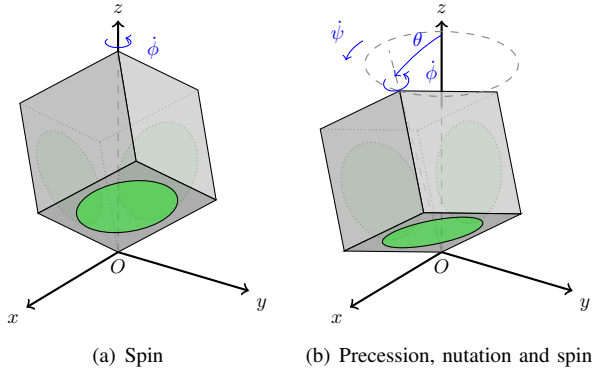

                Cubli has many dynamic equilibrium motions, the most well-known being those similar to the spinning top motion. Two of them will be analyzed: the single spin motion and the precession, nutation and spin motion. The first occurs when Cubli is in its static equilibrium position (either stable or unstable) but spinning around its diagonal axis (Fig. \ref{fig/cubli_dynamic_equilibrium_motions}a). The precession, nutation and spin motion occurs when Cubli center of mass vector $\vec{r}_c$ is not perfectly aligned with the $z$ axis in the inertial coordinate frame, so it spins around its diagonal axis and also precesses around the $z$ axis in the inertial coordinate frame (Fig. \ref{fig/cubli_dynamic_equilibrium_motions}b). For these simulations, the reaction wheels were also assumed to be fixed.
            
                All the simulations were performed utilizing quaternions, but for the easy of representation, the results were then converted to Euler angles.
                
\begin{figure}[H]
                    \centerline{\includegraphics[width=0.5\textwidth]{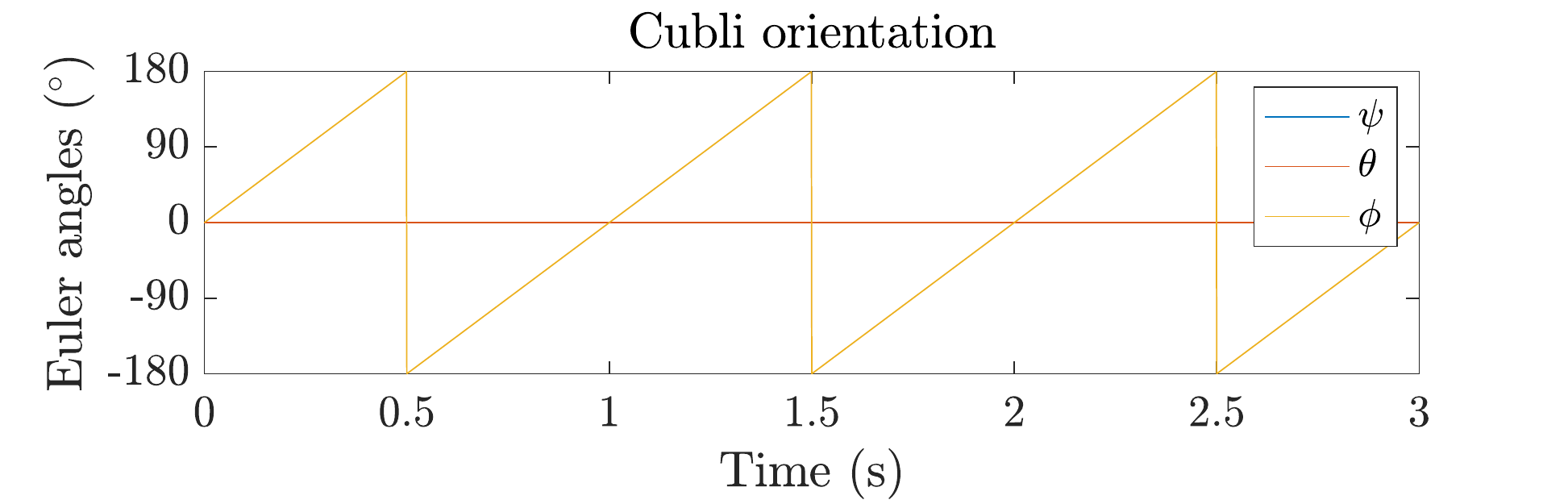}}
                    \caption{Simulation 5 - Dynamic equilibrium (Euler angles)}
                    \label{fig/sim6_singular_motion_dynamic_equilibrium_spin_euler_angles}
                \end{figure}

                For the single spin motion, the same initial conditions as the previous simulation, $q(0)=q_u$, were assumed, but now with $\vec{\omega}_c\,'(0) = {\begin{bmatrix} \frac{2\pi}{\sqrt{3}} & \frac{2\pi}{\sqrt{3}} & \frac{2\pi}{\sqrt{3}}\end{bmatrix}}^T$, meaning it is spinning. Results are presented in Fig. \ref{fig/sim6_singular_motion_dynamic_equilibrium_spin_euler_angles}. 
                
                The spin angle kept increasing while the precession and nutation angles remained unchanged, meaning that Cubli only rotates around its diagonal axis. Moreover, Cubli rotates at exactly $2\pi$ rad/s ($1$Hz), which agrees with the initial angular velocities.
                    
\begin{figure}[H]
                    \centerline{\includegraphics[width=0.5\textwidth]{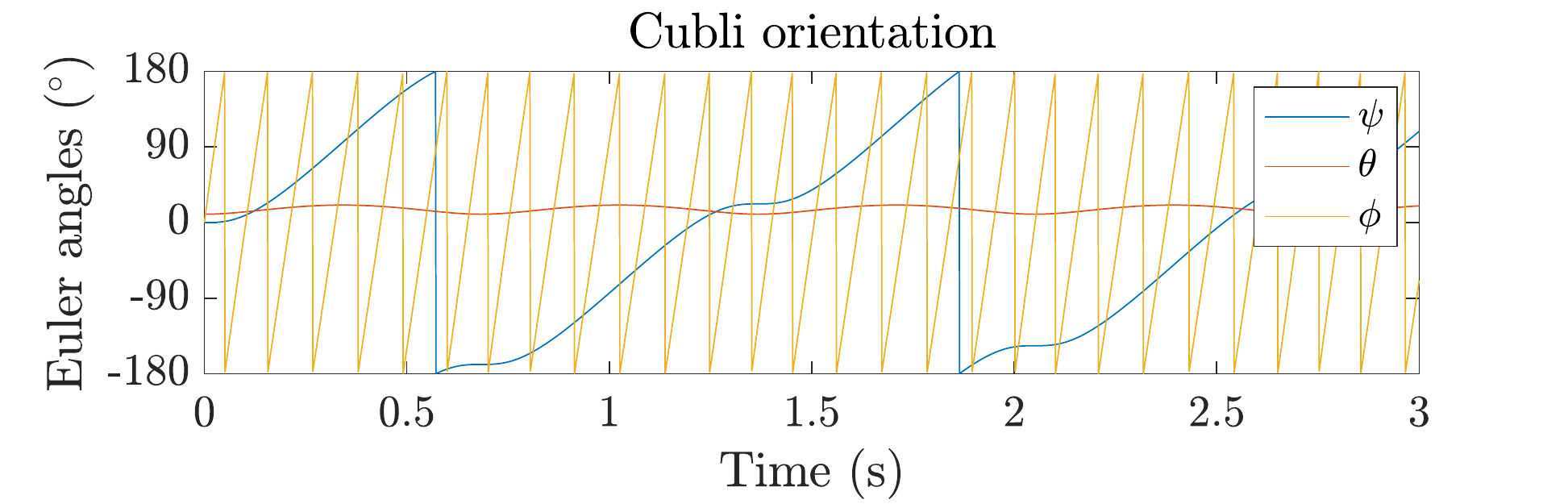}}
                    \caption{Simulation 6 - Dynamic equilibrium (Euler angles)}
                    \label{fig/sim9_singular_motion_dynamic_equilibrium_precession_nutation_spin_euler_angles}
                \end{figure}

                To simulate the precession, nutation and spin motion, a non-equilibrium rotation quaternion $q_{ne}=$\newline${\begin{bmatrix} 0.93 & 0.27 & -0.27 & 0 \end{bmatrix}}^T$ was calculated considering a somewhat small nutation angle ($10^{\circ}$). Considering this new rotation quaternion as an initial condition, $q(0) = q_{ne}$, and with $\vec{\omega}_c\,'(0) ={\begin{bmatrix} \frac{20\pi}{\sqrt{3}} & \frac{20\pi}{\sqrt{3}} & \frac{20\pi}{\sqrt{3}}\end{bmatrix}}^T$ (to guarantee it precesses). Fig. \ref{fig/sim9_singular_motion_dynamic_equilibrium_precession_nutation_spin_euler_angles} presents the results.
                
                Now all the three angles are changing, but in an organized way. The nutation angle keeps oscillating around $10^{\circ}$, while the precession and spin angle kept increasing. Moreover, the spin velocity is higher than the precession velocity, which is, in fact, expected in the spinning top motion. Although the spin velocity is 10 times that of the previous simulation, the frequency is not 10 times higher, meaning that the spin is now somewhat slower. This is because Cubli is now also performing a gyroscopic precession. 
                            
\begin{figure}[H]
                    \centerline{\includegraphics[width=0.4\textwidth]{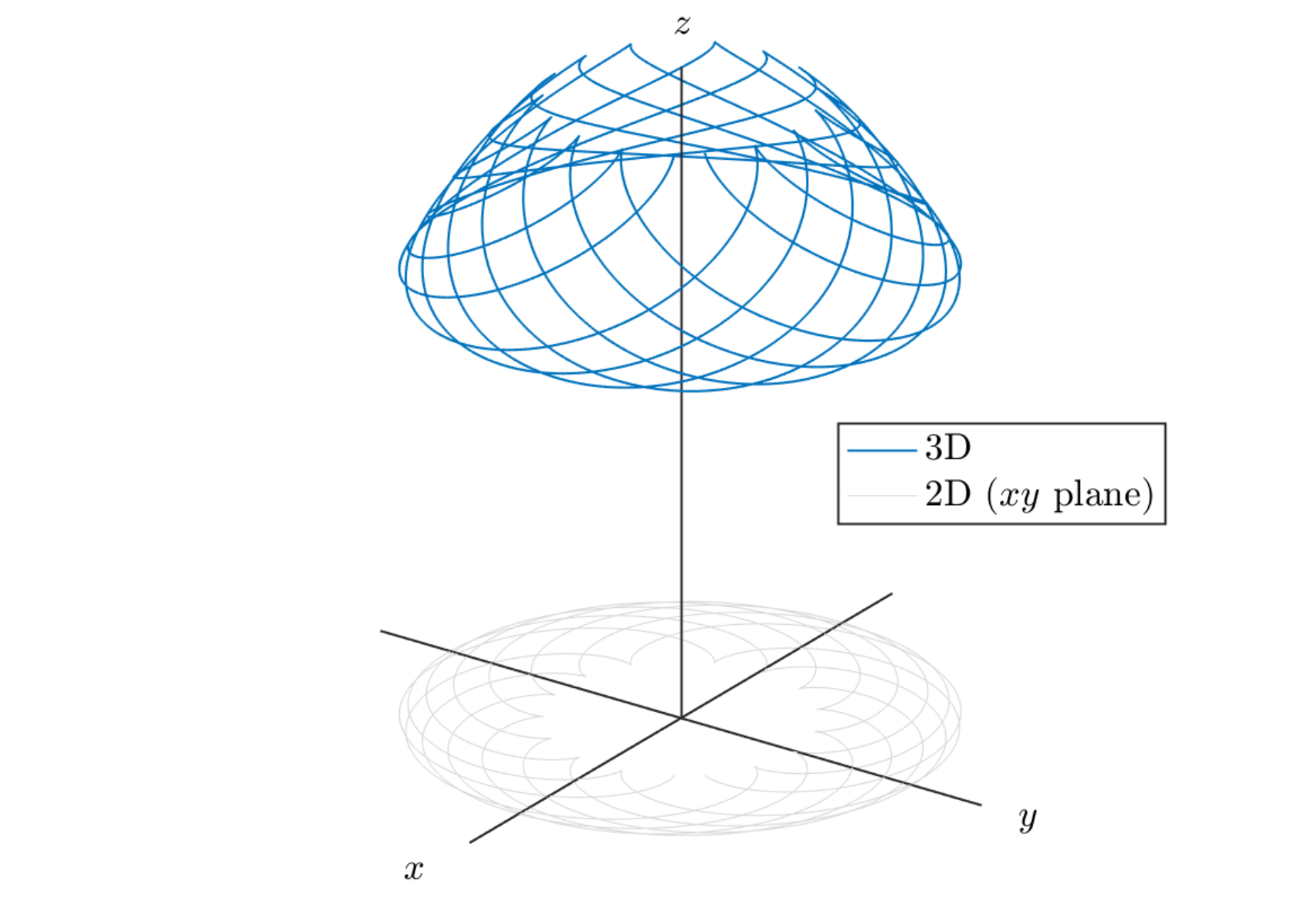}}
                    \caption{Simulation 6 - Dynamic equilibrium (center of mass)}
                    \label{fig/sim9_singular_motion_dynamic_equilibrium_precession_nutation_spin_center_of_mass}
                \end{figure}

                Another interesting graph is the three-dimensional position of Cubli center of mass, which can be seen in Fig. \ref{fig/sim9_singular_motion_dynamic_equilibrium_precession_nutation_spin_center_of_mass}. Although not in scale, it gives a clear perspective of the spinning top motion. 
                
\begin{figure}[H]
                    \centerline{% Set angle of vision
\tdplotsetmaincoords{75}{115}

\begin{tikzpicture}[tdplot_main_coords,scale=0.425]

	% Draw inertial reference frame
	\coordinate[label = below:{$O$}](O) at (0,0,0);
	%\coordinate (O) at (0,0,0) node[anchor=north]{$O$};

	% Draw body reference frame of the back
    \tdplotsetrotatedcoords{0}{0}{0}
	\draw[tdplot_rotated_coords,dashed,red] (0,0,0) -- (4,0,0);
	\draw[tdplot_rotated_coords,dashed,red] (0,0,0) -- (0,4,0);
	\draw[tdplot_rotated_coords,dashed,red] (0,0,0) -- (0,0,4);
	\draw[tdplot_rotated_coords,->,thick,red] (4,0,0) -- (6,0,0) node[below left]{$x\,'$};
	\draw[tdplot_rotated_coords,->,thick,red] (0,4,0) -- (0,6,0) node[right]{$y\,'$};
	\draw[tdplot_rotated_coords,->,thick,red] (0,0,4) -- (0,0,6) node[above]{$z\,'$};
     
    % Draw lines and labels inside cube
    
    \tdplotsetrotatedcoords{0}{0}{0}
    \draw[tdplot_rotated_coords,densely dotted,fill opacity=0.2,fill=green] (2,2,0) circle(1.6);
    \tdplotsetrotatedcoords{0}{90}{90}
    \draw[tdplot_rotated_coords,densely dotted,fill opacity=0.2,fill=green] (2,2,0) circle(1.6);
    \tdplotsetrotatedcoords{-90}{-90}{0}
    \draw[tdplot_rotated_coords,densely dotted,fill opacity=0.2,fill=green] (2,2,0) circle(1.6);
       
    % Draw cube
    \tdplotsetrotatedcoords{0}{0}{0}
    \draw[tdplot_rotated_coords,fill=gray!60,fill opacity=0.80] (0,0,4) -- (4,0,4) -- (4,4,4) -- (0,4,4) -- (0,0,4);
    \draw[tdplot_rotated_coords,fill=gray!40,fill opacity=0.80] (4,0,0) -- (4,4,0) -- (4,4,4) -- (4,0,4) -- (4,0,0);
    \draw[tdplot_rotated_coords,fill=gray!80,fill opacity=0.80] (0,4,0) -- (0,4,4) -- (4,4,4) -- (4,4,0) -- (0,4,0);
    
    % Draw lines and labels outside cube
    \tdplotsetrotatedcoords{45}{54.74}{-90}
	\draw[tdplot_rotated_coords,dashed,purple!50!black] (0,0,0) -- (1.8,0,0);
	\draw[tdplot_rotated_coords,dashed,purple!50!black] (0,0,0) -- (0,1.4,0);
	\draw[tdplot_rotated_coords,dashed,purple!50!black] (0,0,0) -- (0,0,6.93);
    \draw[tdplot_rotated_coords,->,thick,purple!50!black] (1.8,0,0) -- (5,0,0) node[left]{$1$};
    \draw[tdplot_rotated_coords,->,thick,purple!50!black] (0,1.4,0) -- (0,4,0) node[below]{$2$};
    \draw[tdplot_rotated_coords,->,thick,purple!50!black] (0,0,6.93) -- (0,0,12) node[above right]{$3$};

\end{tikzpicture}}
                    \caption{Cubli principal axes}
                    \label{fig/cubli_principal_axes}
                \end{figure}
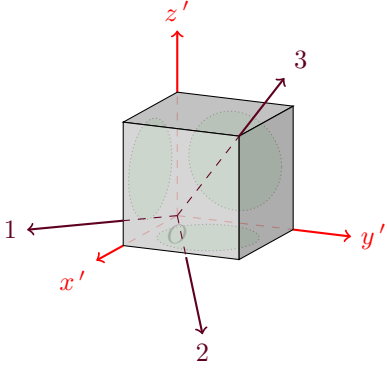

                To completely validate this motion, they will be compared to the well-known \cite{meriam_2012} general equations of rotation of a symmetrical body about a fixed point $O$:
                \begin{equation}
                    \resizebox{0.475\textwidth}{!}{$
                    \left\{
                    \begin{array}{l}
                        I_o \left( \ddot{\psi} \sin\theta + 2 \dot{\psi} \dot{\theta} \cos\theta \right) - I \dot{\theta} \left( \dot{\psi}\cos\theta + \dot{\phi} \right) = 0 \\
                        I_o \left( \ddot{\theta} - \dot{\psi}^2\sin\theta\cos\theta \right) + I \dot{\psi} \left( \dot{\psi}\cos\theta + \dot{\phi} \right) \sin\theta = m g z_G \sin\theta \\
                        I \left( \ddot{\phi} + \ddot{\psi} \cos\theta - \dot{\psi} \dot{\theta} \sin\theta \right) = 0
                    \end{array}
                    \right.
                    $}
                    \label{eqn/spinning_top_dynamics}
                \end{equation}
                where $m$ is the total mass and $z_G$ is the distance from the pivot point $O$ to $G_c$, given by $|\vec{r}_c\,'|$. $I_o$ and $I$ are the moments of inertia around the principal axes, where $I_o = I_{11} = I_{22}$ and $I = I_{33}$ (Fig. \ref{fig/cubli_principal_axes}). The principal axes are obtained from the rotation matrix composed of the eigenvectors of the non-principle axes inertia tensor.
                
                Considering the same previous initial conditions, but now in terms of Euler angles, $\psi(0) = \phi(0) = \dot{\psi}(0) = \dot{\theta}(0) = 0$, $\theta (0) = 10^{\circ}$ and $\dot{\phi} (0) = 30\pi$rad/s, and simulating Eq. (\ref{eqn/spinning_top_dynamics}), the result is the same of Fig. \ref{fig/sim9_singular_motion_dynamic_equilibrium_precession_nutation_spin_euler_angles}, confirming that the dynamic equations are consistent.
                
                Next, the steady precession case is considered. Cubli presents a constant spin and precession velocities, and also a constant nutation angle, meaning that $\ddot{\psi} = \ddot{\phi} = \dot{\theta} = 0$. This simplifies Eq. (\ref{eqn/spinning_top_dynamics}) to a single equation:
                \begin{equation}
                    \left(Io-I\right) \dot{\psi}^2\cos\theta - I \dot{\psi} \dot{\phi} + m g z_G = 0
                    \label{eqn/spinning_top_dynamics_steady_precession}
                \end{equation}
                
                From Eq. (\ref{eqn/spinning_top_dynamics_steady_precession}) it is possible to calculate the precession velocity from the spin velocity and nutation angle:
                \begin{equation}
                    \dot{\psi} = \frac{I\dot{\phi}\pm \sqrt{I^2\dot{\phi}^2-4\left(Io-I\right)\cos\theta mgz_G}}{2\left(Io-I\right)\cos\theta}
                    \label{eqn/precession_velocity}
                \end{equation}
                
                Note that for this equation to be valid, the square root term must be real, meaning that there is a minimum spin velocity needed for steady precession, i.e.:
                \begin{equation}
                    \dot{\phi} \geq \frac{2}{I}\sqrt{\left(Io-I\right)\cos\theta mgz_G}
                    \label{eqn/spin_velocity}
                \end{equation}
                \begin{figure}[H]
                    \centerline{\includegraphics[width=0.5\textwidth]{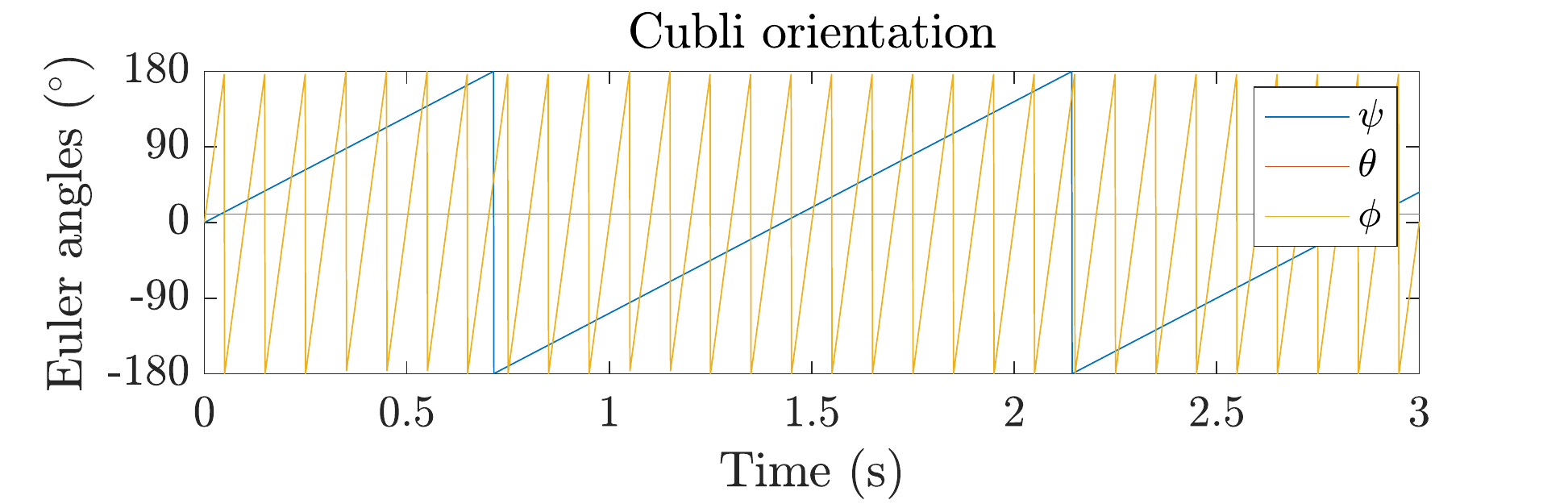}}
                    \caption{Simulation 8 - Dynamic equilibrium (Euler angles)}
                    \label{fig/sim8_singular_motion_dynamic_equilibrium_precession_nutation_spin_euler_angles}
                \end{figure}
                
                The spin velocity and nutation angle satisfy Eq. (\ref{eqn/spin_velocity}). From (\ref{eqn/precession_velocity}) it is possible to see that $\psi_1(0) = 4.40$ rad/s or $\psi_2(0) = 22.19$ rad/s. Writing one of this initial angular velocities in terms of Cubli dynamics, since we want to simulate  (\ref{eqn/cubli_dynamic_equations}) instead of (\ref{eqn/spinning_top_dynamics}), it means that $\vec{\omega}_c\,'(0) = {\begin{bmatrix} 38.47 & 38.47 & 39.40 \end{bmatrix}}^T$. Results are presented in Fig. \ref{fig/sim8_singular_motion_dynamic_equilibrium_precession_nutation_spin_euler_angles}, where Cubli is now clearly in steady precession, showing once again the model consistency.

        \subsection{Poinsot trajectories}
        \label{sec/poinsot_trajectories}
    
            Poinsot trajectories are a geometrical method for visualizing the torque-free motion of a rotating rigid body. Since the system needs to be in torque-free motion, gravity will be neglected. The law of angular momentum conservation implies that in the absence of applied torques,  $\vec{H}$ is conserved in an inertial reference frame ($\frac{d\vec{H}}{dt} = 0$). The law of energy conservation implies that in the absence of input torques and energy dissipation, $T$ is also conserved ($\frac{dT}{dt} = 0$).  Considering the principal axes, it is possible to write $\vec{H} = \begin{bmatrix} I_o \omega_1 & I_o \omega_2 & I \omega_3 \end{bmatrix}^T$, so that the total angular momentum is simply the magnitude of this vector:
            \begin{equation}
                H = \sqrt{I_o^2 \omega_1^2 + I_o^2 \omega_2^2 + I^2 \omega_3^2}
                \label{eqn/poinsot_angular_momentum}
            \end{equation}
            
            The angular kinetic energy, also considering the principal axes, is given by:
            \begin{equation}
                T = \frac{1}{2} I_o \omega_1^2 + \frac{1}{2} I_o \omega_2^2 + \frac{1}{2} I \omega_3^2
                \label{eqn/poinsot_kinetic_energy}
            \end{equation}
            
            Writing (\ref{eqn/poinsot_angular_momentum}) and (\ref{eqn/poinsot_kinetic_energy}) in terms of the components of the angular momentum vector along the principal axes: 
            \begin{equation}
                \left\{
                \begin{array}{l}
                    H^2 = H_1^2 + H_2^2 + H_3^2 \\
                    2T = \frac{H_1^2}{I_o} + \frac{H_2^2}{I_o} + \frac{H_3^2}{I} 
                \end{array}
                \right.,
            \end{equation}
            which are equivalent to two constraints for the 3D angular momentum vector $\vec{H}$. The angular momentum constrains $\vec{H}$ lie on a sphere, whereas the kinetic energy constrains $\vec{H}$ lie on an ellipsoid. These two surfaces intersections define the possible solutions for $\vec{H}$.
            
\begin{figure}[hbt!]
	\centering
	\subfigure[][]{\includegraphics[trim=2.1cm 0cm 2.55cm 0cm, clip=true, width=.45\columnwidth]{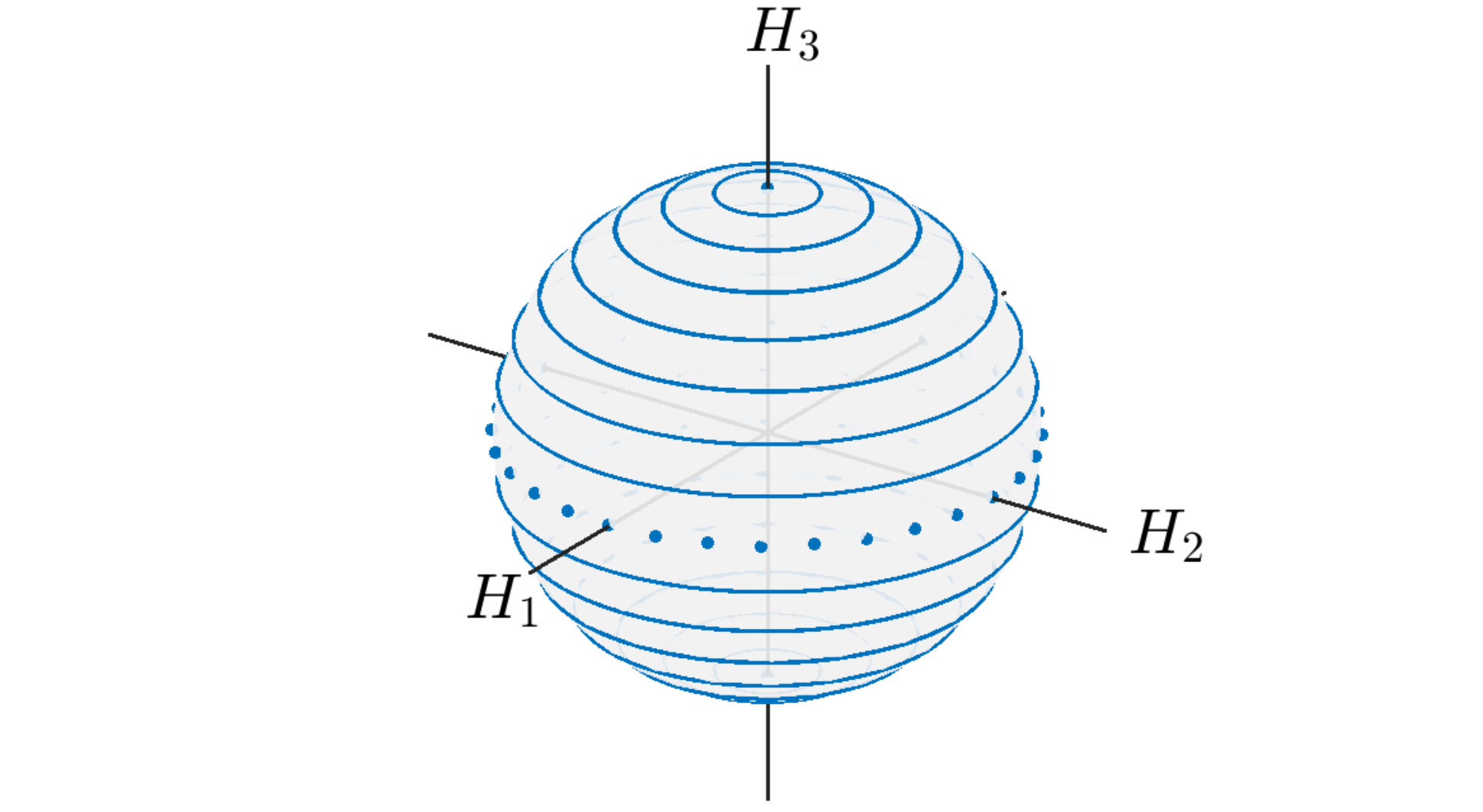}\label{fig/sim9_poinsot_trajectories_kinetic_energy}}
	\subfigure[][]{\includegraphics[trim=2.1cm 0cm 2.55cm 0cm, clip=true, width=.45\columnwidth]{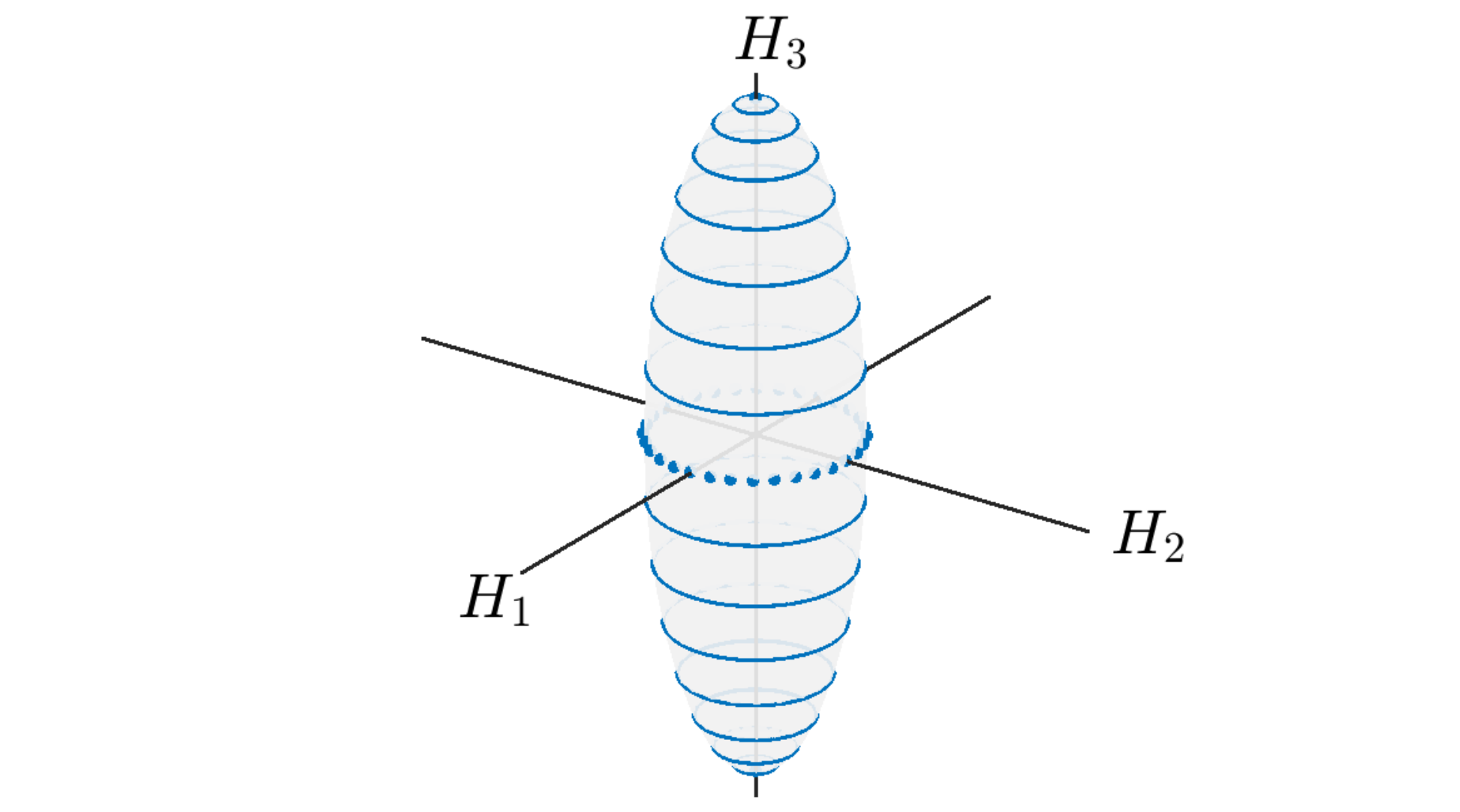}\label{fig/sim9_poinsot_trajectories_angular_momentum}}
	\caption{Simulation 9 -  (a) Poinsot traj. (constant angular momentum); (b) Poinsot traj. (constant kinetic energy) }
\end{figure}

            Simulations in Fig. \ref{fig/sim9_poinsot_trajectories_kinetic_energy} considered various values of initial angular velocities, all with the same total $H$.

The surface is a sphere, which is expected for a constant angular momentum. Each line or dot is a different simulation and represents an intersection with the kinetic energy ellipsoid. Moreover, because each simulation has constant $H_3$, the body is axisymmetric through this axis, which is in fact the case for Cubli. Considering now the same kinetic energy $T$, yields Fig.  \ref{fig/sim9_poinsot_trajectories_angular_momentum}. In this case, the surface is an ellipsoid, which is also expected for a constant kinetic energy. Now each line or dot represents an intersection with the angular momentum sphere. Moreover, because two moments of inertia are the same and the third one is smaller than the other two, the shape is in fact a prolate spheroid, which is the particular case of an ellipsoid. 

    \section{Conclusions}
    
        By utilizing quaternions instead of Euler angles, modeling could be performed utilizing vector notation. Although a bit complex in the beginning (given its different algebra), at the end the dynamic equations were quite compact and obtained completely by hand, without the need of any mathematical symbolic software. 
        
        Computer simulations showed that the model is consistent, while Poinsot trajectories presented a geometrical approach that also validated the model.
        
        As for future works, the control design utilizing quaternions is under development, as well as the implementation of the system in a real-world prototype.
    %\nocite{*}
    \bibliographystyle{ieeetr}
    \bibliography{biblio}

\end{document}